# Emission and Propagation of Multi-Dimensional Spin Waves with nanoscale wavelengths in Anisotropic Spin Textures

V. Sluka[1,a,*], T. Schneider[1], R. A. Gallardo[2,3], A. Kakay[1], M. Weigand[4], T. Warnatz[1,b], R. Mattheis[5], A. Roldan-Molina[6], P. Landeros[2,3], V. Tiberkevich[7], A. Slavin[7], G. Schütz[4], A. Erbe[1], A. Deac[1], J. Lindner[1], J. Raabe[8], J. Fassbender[1,9], and S. Wintz[1,8,*]

**Spin waves offer intriguing novel perspectives for computing and signal processing, since their damping can be lower than the Ohmic losses in conventional CMOS circuits. For controlling the spatial extent and propagation of spin waves on the actual chip, magnetic domain walls show considerable potential as magnonic waveguides. However, low-loss guidance of spin waves with nanoscale wavelengths, in particular around angled tracks, remains to be shown. Here we experimentally demonstrate that such advanced control of propagating spin waves can be obtained using natural features of magnetic order in an interlayer exchange-coupled, anisotropic ferromagnetic bilayer. Using Scanning Transmission X-Ray Microscopy, we image generation of spin waves and their propagation across distances exceeding multiple times the wavelength, in extended planar geometries as well as along one-dimensional domain walls, which can be straight and curved. The observed range of wavelengths is between 1 µm and 150 nm, at corresponding excitation frequencies from 250 MHz to 3 GHz. Our results show routes towards practical implementation of magnonic waveguides employing domain walls in future spin wave logic and computational circuits.**

[1] Helmholtz-Zentrum Dresden-Rossendorf, 01328 Dresden, Germany
[2] Universidad Técnica Federico Santa María, 2390123 Valparaíso, Chile
[3] Center for the Development of Nanoscience and Nanotechnology (CEDENNA), 917-0124 Santiago, Chile
[4] Max-Planck-Institut für Intelligente Systeme, 70569 Stuttgart, Germany
[5] Leibniz Institut für Photonische Technologien, 07745 Jena, Germany
[6] Universidad de Aysén, Calle Obispo Vielmo 62, Coyhaique, Chile
[7] Oakland University, Rochester, MI 48309, USA
[8] Paul Scherrer Institut, 5232 Villigen PSI, Switzerland
[9] Technische Universität Dresden, 01069 Dresden, Germany
[a] Present address: New York University, New York, NY 10003, USA
[b] Present address: Uppsala Universitet, 75120 Uppsala, Sweden
[*] e-mail: vs1568@nyu.edu; s.wintz@hzdr.de

Spin waves are the elementary excitations of the order parameter in ferromagnetic materials [Fig. 1a)] [1-3]. Also referred to as magnons, they can be used similarly to electrons in CMOS circuitry, but with lower losses, to transmit information, and, therefore, are currently attracting a lot of interest as possible information carriers in alternative computing schemes [4-7]. One of the most pressing issues in present-day high-performance computing are the high power requirements and the necessary heat removal associated with the Ohmic losses in conventional electronic CMOS circuits - latest generation supercomputers easily consume power in the order of ten Megawatts. On a wider societal scale, the reduction of signal-processing losses, in particular in personal mobile communication devices may create substantial benefits due to reduced power consumption, resulting in extended battery life and improved environmental sustainability. Another substantial advantage of spin-wave technology is the fact, that in the GHz range, magnon wavelengths are several orders of magnitude shorter than those of electromagnetic waves [8]. Thus, a significant device miniaturisation can be achieved for applications where the wavelength imposes a critical constraint on the device footprint. For such purposes it will be crucial to utilise spin waves with wavelengths in the sub-µm range, where both magnetostatic and exchange effects are relevant (dipole-exchange waves) [5]. While surface acoustic waves are already present as short-wavelength signal carrier in today's communication technology, spin waves offer a superior scalability of wave excitation and propagation at frequencies above 2 GHz as well as a much wider frequency tunability [9,10].

Two of the most challenging aspects of building a magnonic computer remain the generation of short-wavelength magnons and the construction of suitable waveguides for spin wave transport. Several recent works focus on these two issues ([11-29] and [7,29-40], respectively). A standard method to coherently generate spin waves employs the localised Oersted fields from alternating electric currents flowing in metallic antennas that are patterned adjacent to a magnetic medium. The smallest excitable wavelengths using this method, however, are approximately equal to the patterning sizes involved [cf. Supplemental information (SI) (3)]. In terms of nanopatterning and microwave impedance matching, it is therefore highly challenging to efficiently scale such an antenna-based excitation to nanoscale wavelengths. While similar restrictions apply from the patterning size, spin-transfer torques have proven to be an alternative suitable source for spin-wave excitation [13,14,16,26], with the possibility for steering spin waves by external magnetic bias fields [26]. More recently, it was also shown that spin waves can be generated using the internal fields of non-uniform spin textures [15,17-25,27-29], yet direct observations are limited to wavelengths > 1μm, apart from [18,29]. In [18] the emission of high-amplitude nanoscale spin waves from a pair of stacked vortex cores [41], driven by an alternating magnetic field [42,43] was demonstrated. However, in the geometry [18], where a point-like vortex core source is radiating spin waves into a two-dimensional propagation medium, spin waves originating from a vortex core and traveling outwards radially experience not only Gilbert damping, but also a purely geometric reduction of amplitude proportional to the inverse square root of the distance from the source, as shown schematically in Fig. 1b). Further, it has been suggested that magnetic domain walls could be harnessed to guide spin waves across the magnonic chip [7,29-36]. In particular, it has been shown that domain walls can host localized modes, excited by alternating magnetic fields [31]. While in Ref. 31, the lateral position of the excited magnetisation amplitudes could be well controlled by tuning the lateral domain wall position, these modes however quickly decayed along the domain wall coordinate with increasing distance from the microwave antenna within subwavelength length scales.

These key issues – short wavelength-spin wave generation and spin wave guidance - are the points we address in this work, where we make use of naturally formed anisotropic spin textures. First, we demonstrate the excitation and propagation of two-dimensional planar spin-waves [Fig. 1c)] excited by the oscillation of straight domain walls. We observe that these excitations can travel distances spanning multiples of the wavelength. Second, we observe excitation and propagation of spin wave modes confined to quasi-one-dimensional natural waveguides (straight or curved) formed by domain walls embedded in a two-dimensional host medium [Fig. 1d)].

Our samples are $Co_{40}Fe_{40}B_{20}/Ru/Ni_{81}Fe_{19}$ multilayers with (46.6/0.8/44.9) nm thickness, patterned into disc- and square-shaped elements, having lateral sizes of several microns [Fig. 2a)]. Each ferromagnetic layer exhibits an in-plane uniaxial anisotropy. The Ru interlayer causes antiferromagnetic coupling between the two ferromagnetic layers [44] (see methods part for further details).

The magnetic ground state configuration stabilized in this system is a pair of stacked vortices, with opposite vorticity due to the antiferromagnetic interlayer exchange coupling. The influence of the CoFeB uniaxial anisotropy leads to a significant distortion of the vortex magnetization distribution in both magnetic layers. The result in each layer is a state of two homogeneously in-plane magnetized domains with opposite magnetizations. These domains are separated by a narrow, partially perpendicularly oriented, 180 degree domain wall that contains the vortex cores and spans the lateral extension of the discs. These magnetic configurations are shown in Figures 2b) and 2c), which are Scanning Transmission X-ray Microscopy (STXM) images displaying magnetic information about the in-plane (b) and the out-of plane component (c) of the individual layers, where the technique provides a lateral resolution of about 25 nm. As Fig. 2c) indicates, the out-of-plane magnetization components of the respective layers couple ferromagnetically to each other by their stray field. In particular, this is true for the polarizations of the vortex cores. Micromagnetic simulations confirm this and reveal that the domain wall formed in the sample is, in fact, a mixture between Néel and Bloch types of domain walls [45], where the in-plane components couple antiferromagnetically across the Ru interlayer, as in the domains. The complex ground state magnetic pattern is illustrated in panels d) and e) of Fig. 2. Figure 2d) displays a schematic top view of the domain wall structure in the CoFeB layer, showing the mixed Bloch and Néel components. In Fig. 2e) a cross-section of the bilayer system is shown, which can be imagined as resulting from a cut along the blue lines in panel c) of the Figure, revealing the out-of-plane magnetization components in the domain wall to follow a flux-closing distribution between the two layers.

Spin waves can be excited in such anisotropic spin textures by applying an alternating magnetic field, as shown in Figure 3. The corresponding measurements were made by means of time-resolved (TR)-STXM imaging, allowing for a stroboscopic time-resolution of approximately 100 ps. Figure 3 (a) is a snapshot of the magnetic excitations at an Oersted field frequency of 1.11 GHz, taken at the Ni absorption edge, displaying the out-of-plane contrast. Plane spin waves are visible, with wave fronts parallel to the domain wall, and propagating away from the domain wall towards the rim of the elliptical element, as indicated by the green arrow. The oscillating Oersted field in-plane component is oriented along the minor axis of the ellipse, perpendicular to the domain wall. The main effect of the Oersted field is to excite dynamics of the domain wall, and that the excited domain wall acts as a confined perpendicular source for the observed spin waves [24,25,29]. In more detail, by acting on the full sample volume, the field excites a non-resonant antiphase width oscillation of the walls in the two different layers which is coherent over the wall length, causing highly localised out-of-plane torques in the wall vicinity. Thereby spin waves are essentially excited via a linear and coherent coupling of the discrete wall mode to the spin-wave continuum, similar to the situation of vortex core driven spin waves reported earlier [18,27].

The periodic-in-time nature of the waves allows capturing the wave motion at discrete, equispaced phases in each scanned pixel, and composing the recorded data into movie-like arrangements, which impressively show the propagation of these spin-excitations [see movies in the SI (1)]. A comparison of the absorption data taken along the green arrow in Fig. 3a) at different time slices yields the wavelength of the wave, and in particular its speed of propagation. Three of these time slices are shown in Fig. 3b). Notably, the spin wave amplitude does not visibly decrease across the distance of two micrometers, corresponding to about 7.5 times the wavelength. Increasing the excitation frequency to 1.46 GHz results in a similar wave pattern, but with shorter wavelength [Fig. 3c)]. By comparison with the contrast of the vortex core, the spin-wave amplitudes are estimated to reach a precession angle of beyond 5 degrees, which can be considered as very high when compared to standard spin-wave excitation techniques.

Around the vortex center, as shown in the magnified image Fig. 3d), in addition to the plane waves generated by the oscillating wall, there exist radial wave fronts which arise from the motion of the vortex core, which acts as a point source [18]. In comparison to the plane waves excited by the domain wall, these radially symmetric waves must decay faster in power density, with a factor of $1/r$ in addition to the exponential decay induced by the Gilbert damping, $r$ corresponding to the distance from the vortex center. This difference can be regarded as a consequence of the fact that the plane waves are excited by a one-dimensional source (the domain wall), while in the case of the radial waves, the source is zero-dimensional (the vortex core). As these two wave forms are excited simultaneously, patterns of interference arise which are also visible in Fig. 3d).

In this manner, we can excite planar spin waves for a broad range of frequencies up to 3 GHz. Yet, one can expect this process to scale to even higher frequencies, if the magnetisation gradient of the exciting source (domain wall or vortex core) was enhanced as for domain walls in systems with strong perpendicular magnetic anisotropy or if the spin-wave dispersion relation was tuned to longer wavelengths by modifying the magnetic layer stack. Remarkable effects, however, appear when going to rather low excitation frequencies, as shown in Fig. 4, displaying excitations at 0.52 GHz and 0.26 GHz [panel a) and b), respectively]. At these low frequencies, no visible excitations exist in the domains, yet the data clearly shows spin waves propagating confined to the domain wall in the directions away from the vortex cores. The wavelength of these waves can be controlled in the same way as in the previous cases, *i.e.* by tuning the excitation frequency. Note that there is a directional asymmetry in the spin-wave emission at

0.52 GHz, which, however can be attributed to sample imperfections in the exciting core region as simulations indicate a symmetric emission pattern. Irrespective of this, the wave amplitude is again still significant even after a propagation distance extending from the vortex core to the rim of the ellipse. This is made possible in a way that is analogue to the above described case of planar waves in a two-dimensional medium excited by the one-dimensional domain wall: For the waves propagating along the domain wall, the source is of dimension zero; however, due to the confinement to the domain wall, the propagation medium is effectively one-dimensional. As a result, geometrical decay of the amplitude is avoided, making the domain wall act as a low-loss waveguide (c.f. SI movie M4 of propagating spin waves in the domain walls). The confined waves excited here can be considered a bilayer analog of the spin-wave mode predicted by *Winter* for a single domain wall [33], while also the coherent domain wall resonance (or infinite wavelength $k$=0) case of such a bilayer has been theoretically studied in Ref. [46] [cf. SI(2)].

In order to shed light on the physics underlying these observations, we followed a twofold strategy: First, both observed phenomena - the excitation and propagation of planar spin waves in the domain, and one-dimensional waves confined to the domain wall – were investigated and qualitatively confirmed with micromagnetic simulations. For that purpose, the experimental static magnetization distribution was reproduced prior to excitation by an ac magnetic field. We refer to SI movie M7 for a qualitative confirmation of the gapless one-dimensional spin-wave mode in the domain wall. In order to obtain the details of the dispersion relation for the planar waves in the domains within reasonable computation time, the system was modeled by two continuous, homogeneously magnetized coupled layers (further details can be found in the methods section). The experimental plane spin-wave dispersion was quantitatively reproduced by these simulations.

In addition to the simulations, we developed a theory (see SI for in-depth technical details) for the propagation of spin waves in two exchange-coupled extended ferromagnetic films. The core of the theory considers spin wave modes in thin magnetic films, where the magnetization along the coordinate perpendicular to the film plane can be considered homogeneous. The case of thicker films as in the experiment is accounted for by splitting each ferromagnetic layer into a number $N$ of thin films of equal thickness, so that for each of these films the thin-film approximation holds. The $N$ thin films of each layer are then coupled to each other by an effective ferromagnetic intralayer exchange coupling, whose strength is determined by estimating the energy of a magnetization distribution subject to homogeneous torsion and by requiring consistency with the continuum limit. The theory thus enables us to quickly compute dispersion relations for spin waves in the interlayer exchange-coupled bilayer system with ferromagnetic layers whose thicknesses exceed the exchange lengths of the respective material. In Fig. 5, the measured spin wave dispersion relations $f(k = 2\pi/\lambda)$ ($f$ denoting the frequency, $k$ the wave number and $\lambda$ the wavelength, respectively) for waves in the domains and in the domain walls are combined with the analytical and micromagnetic simulation results. We first consider the planar waves propagating through the domains. The open circles represent the results extracted from the STXM measurements. The blue continuous line displays the theoretical result, which is found to depend sensitively on the CoFeB in-plane uniaxial anisotropy and the interlayer exchange coupling. For $J$ = -0.1 mJ/m$^2$ and $K_u$(CoFeB) = 3 kJ/m$^3$, we find good agreement with the experimental data. The elevated value of $K_u$(CoFeB) is reasonable, since we expect the CoFeB to react sensitively to strain exerted by the patterned waveguide microstructure onto the elliptical element [47]. Using the same parameters as in the theory, we also compute the dispersion using micromagnetic simulations [grey dots in Fig. 5)]. Note, that the antiparallel bilayer system considered here can in principle host two separate spin-wave modes of acoustic and optical character [cf. SI (2)]. Since the optical mode however resides at much higher frequencies and is not accessed by our experiments, we only consider the acoustic mode in the following.

A striking feature of the acoustic plane wave dispersion is the existence of a local minimum at low $k$ around 5 rad/μm, and accordingly, a frequency gap, below which no spin wave excitations are possible. The local minimum at some finite value of the wave vector in Fig. 5 can be understood due to a combination of the non-reciprocity induced by the dipolar coupling between the two antiferromagnetically coupled magnetic layers [18,48] [SI (2)] and the uniaxial magnetic anisotropy. Namely, when the anisotropy is null, the collective dispersion in Damon-Eshbach geometry ($\mathbf{k} \perp \mathbf{M_0}$) [49] has a minimum of zero frequency at $k = 0$, while at finite anisotropies, this minimum is shifted to finite values of both wave vector and frequency. Such a $k$-shifting of the dispersion minimum is somewhat analogous to that induced by the Dzyaloshinskii-Moriya interaction on ferromagnetic/heavy-metal alloys, where the minimum of the dispersion is also shifted. [50]. Note that we only observe the slow branch ($k \geqslant +5$ rad/μm) of the non-reciprocal dispersion relation in our experiment since the wavelengths of the fast branch are of the order or exceeding the sample size for the frequencies given [cf. SI (2)]. At the same time the spin-wave amplitudes predicted for the fast branch are also much lower, which both are the reasons for why spin-wave edge reflections are not noticeable in the experiment.

Our experimental observations of selective excitation and propagation of spin waves in the domain wall can actually be explained based on the existence of the frequency gap discussed above: The red circles in Fig. 5 display the dispersion relation of the measured spin waves in the domain wall. In sharp contrast to the planar waves in the domains, the waves confined to the wall exhibit an almost linear dispersion, which runs below its plane wave counterpart and, when extrapolated towards zero, intercepts the y-axis close to $f = 0$. Thus, when tuning the excitation frequency to values inside the gap, no propagating magnons are excited in the domains; only the energetically lower modes existing in the wall are populated. The existence or, respectively, absence of the gaps in the domain and domain wall can be explained by the Goldstone theorem [51], which states that a system exhibiting a continuous symmetry spontaneously broken by the ground state has a gapless mode. In case of the spin waves in the domains, the corresponding system comprises the two coupled discs. Here, the continuous symmetry is compromised by the uniaxial anisotropy and accordingly, the planar spin wave dispersion relation exhibits a gap. In case of the waves confined to the domain wall, there exists a continuous translational symmetry that gives rise to a gapless mode. The presence of defects and the finite size of the sample, in principle, break this symmetry, but the resulting gap is too small to change the quality of the observed effects.

The idea of using domain walls as waveguides is intriguing, and Fig. 6 shows that the above described phenomenon indeed extends to cases where the walls are curved, i.e., lead 'around the corner'. Neither is the concept restricted to continuous wave excitation. Figure 6a) displays the static magnetization configuration (out-of-plane contrast) of a domain wall, apparently of the same type as in the aforementioned cases, but curved towards the right-hand rim of the magnetic element. The regions marked in orange and yellow inside the domain wall indicate positions in front of and behind the curve, respectively, when following the domain wall from the vortex core towards the rim. Panels (b)-(d) display the snapshots of the excitation following a field pulse: Due to the width of the spectral composition of the pulse, spin waves are excited inside and above the frequency gap. The resulting plane wave packet traverses the domains in the direction away from the wall and makes it easy to optically distinguish the domain wall wave from the rest of the excitations. Fig. 6b) displays the time slice just before the field pulse. 11.1 ns after the pulse, the wall wave packet has reached the orange region in front of the turn [Fig. 6c)]; 2.5 ns later the wave packet has traveled around the corner. Remarkably, even after the turn, the wave packet maintains a considerable amplitude. While the spatial distribution of the domain wall in our experiment is solely a consequence of both dipolar sample confinement fields and magnetostrictive anisotropies, it was shown that further control of the domain wall position can be achieved e.g. by exchange bias patterning [34,39], ferroelectric coupling [21,22], or external magnetic fields [31].

To summarize, the work presented here addresses several key aspects of magnonic computing by exploiting magnetic anisotropy. The first aspect is related to energy and signal range. We demonstrated that textures in a magnetization distribution, like domain walls and vortex cores, can serve as sources for the generation of short-wavelength, dipole-exchange spin waves of directional nature, that is, planar waves in magnetic domains and waves confined to domain walls, which due to their geometry, are not subject to the reduction of amplitude due to the geometric dilution of the energy flow. This is an important result, since such waves minimize the losses occurring during propagation. Indeed, we found that the resulting excitations can travel distances easily spanning several microns, *i.e.*, significantly exceeding multiples of the nanoscale wavelengths – a necessary condition, *e.g.*, for magnon interference-based applications. The second aspect is to identify possible waveguides for magnonic chips. Here, we showed that domain walls can serve as such waveguides, combining several useful properties. First, due to their inherent symmetry, and consequently their near-gapless dispersion relation, spin waves can selectively be excited in these structures. In addition, we showed that spin wave packets can travel along angled domain walls while largely maintaining their amplitude. Such possibility of angled signal guidance is vital for chip design, and therefore our result may enable new solutions to the development of magnonic circuits.

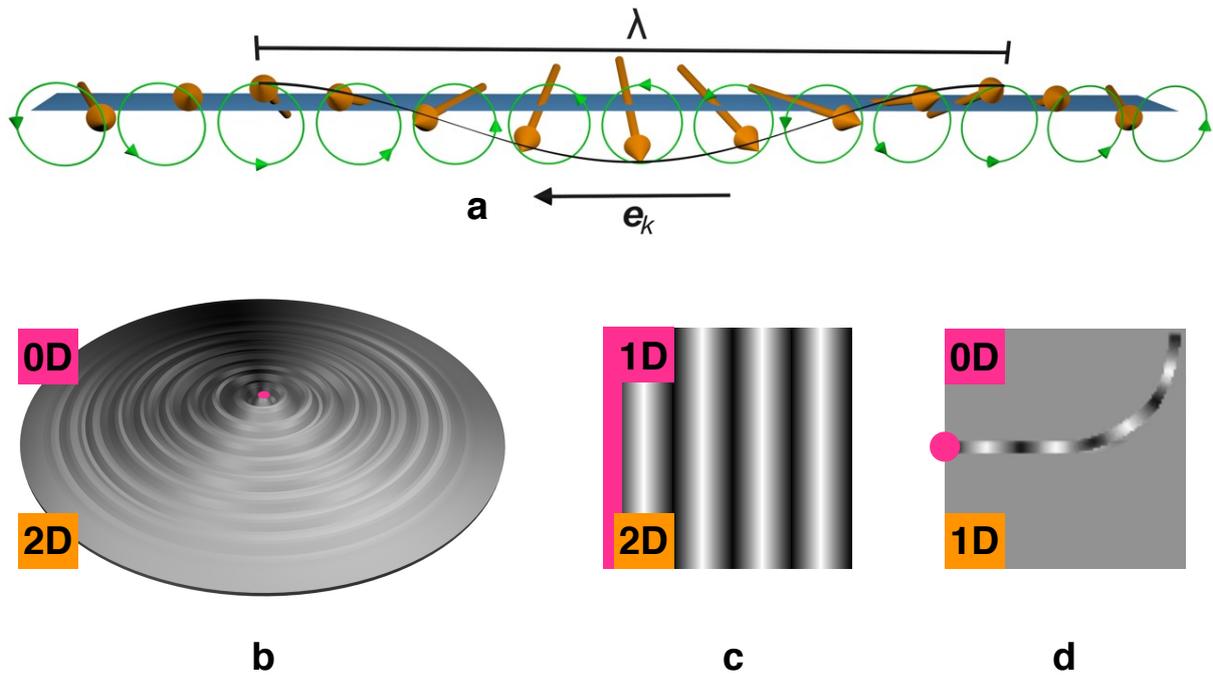

**Figure 1: Spin waves in different geometries. (a)** Schematics of a spin wave propagating along $\mathbf{e}_k$. Magnetic moments (orange arrows) precess with a spatial phase difference, determining the wavelength λ. **(b-d)** Three different geometries of spin wave propagation explored in this paper. The magenta and orange fields denote the geometric dimensions of source and propagation medium, respectively. **(b)** Spin wave emission from a point source. In this case, the dimensions of medium and source differ by two. As a result, in addition to the exponential decay caused by the Gilbert damping, there is a geometric decay of the spin wave amplitude. **(c)** Plane-wave-like spin wave propagation. Similarly to **(b)**, the waveguide medium is two dimensional, however the source has dimension one in this case. Thus, in **c)** the dimensions of source and medium differ by one. This is also the case in panel **(d)**, where a zero-dimensional source excites a one dimensional medium (a domain wall). The situations depicted in **(c)** and **(d)** are of special interest from the viewpoint of engineering magnonic waveguides, since in these cases the losses are largely limited to the Gilbert damping.

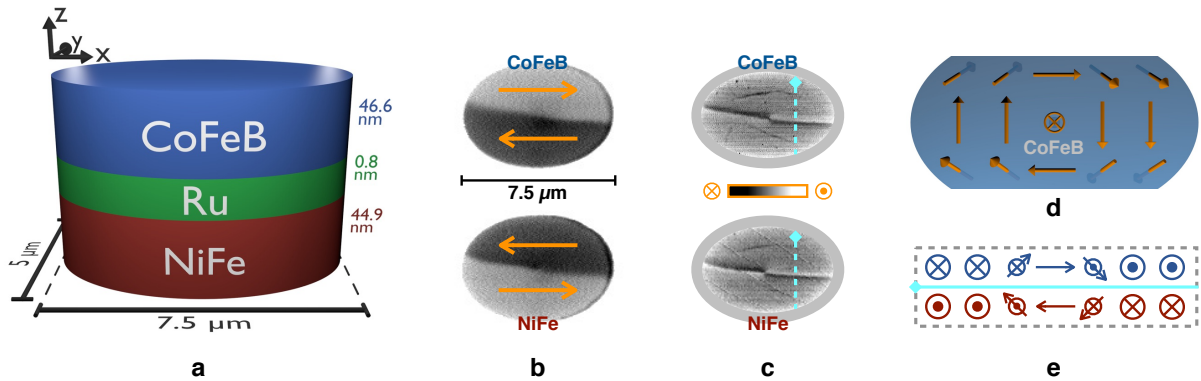

**Figure 2: Sample layout and magnetic configuration. (a)**: The ferromagnetic element is patterned out of an interlayer-exchange-coupled bilayer system, consisting of a NiFe and a CoFeB layer, coupled antiferromagnetically by a Ru interlayer. The coupling leads to the magnetic states shown in **(b)**, where the contrast represents the in-plane component along the long axis of the elliptic element, and **(c)**, where the out-of-plane magnetic contrast is displayed. High-resolution XMCD STXM images show that the magnetic configuration is a pair of stacked vortices, with antiferromagnetically coupled in-plane magnetizations. We find that an additional anisotropy with easy axis along the long axis of the elliptic element leads to an anisotropic deformation of the vortex patterns, resulting in the formation of a domain wall, which is also visible both in the in-plane and out-of-plane contrast images. Micromagnetic simulations reveal that this domain wall has both Neel and Bloch character, as shown in **(d)**. A cross-section of the domain wall profile is shown in **(e)**, which can be imagined as taken along the blue lines in panel **(c)**, illustrating the in- and out-of-plane components of the layer magnetizations in and around the domain wall.

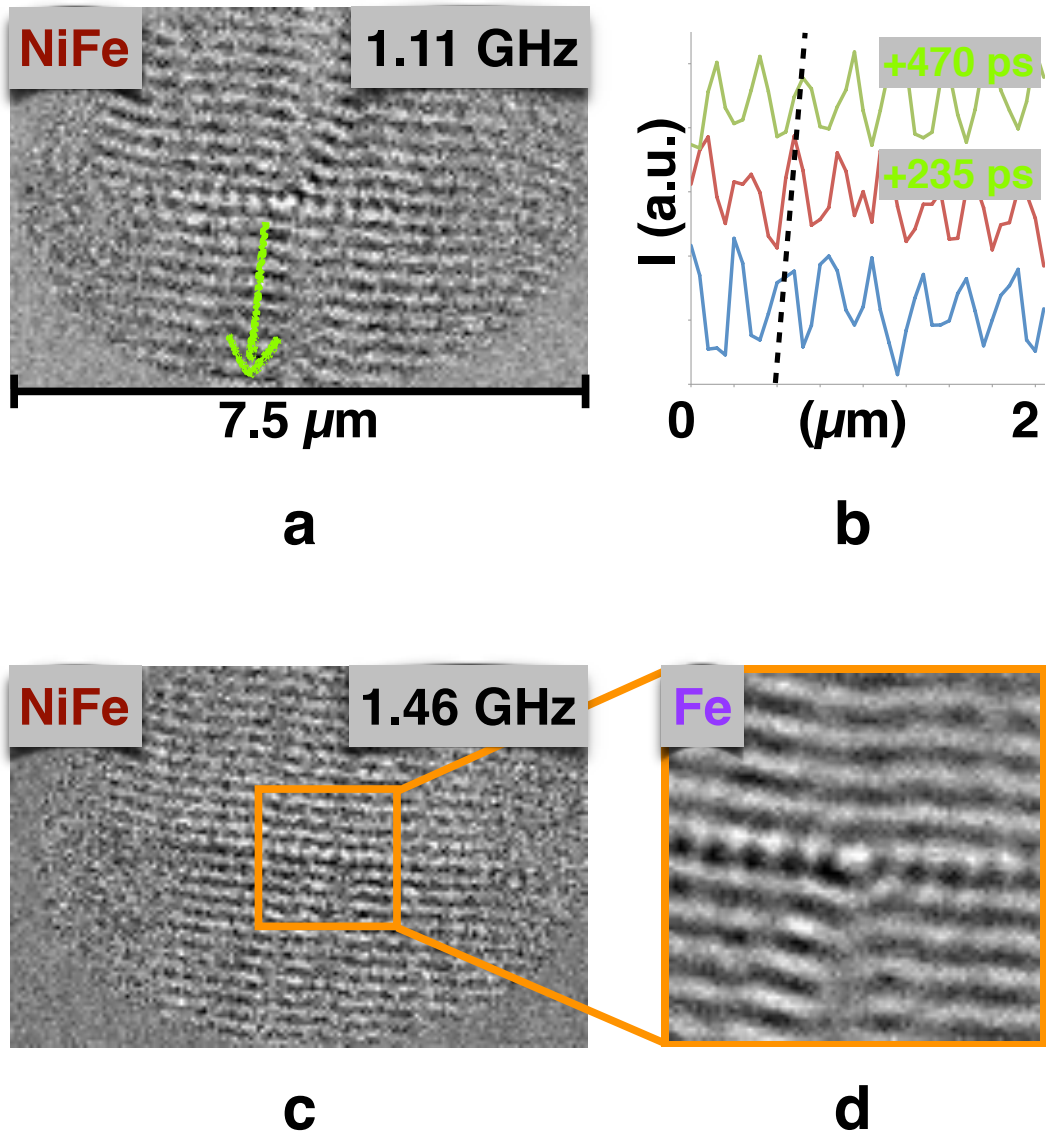

**Figure 3: Excitation of spin waves.** TR-STXM Snapshots of spin waves (NiFe layer out-of-plane magnetization component), excited using ac magnetic fields at different frequencies. **(a)** At an excitation frequency of 1.11 GHz, spin waves are generated that originate from the domain wall spanning the elliptically shaped magnetic element along the long axis through the vortex core. These plane waves travel from the wall to the rim of the disc, as indicated by the green arrow. Three time slices of the signal amplitude along that arrow taken at equidistant time intervals of 235 ps are shown in panel **(b)**. The time slices allow to determine the wavelength or wave vector, respectively. Furthermore, they show that the wave amplitude does not change significantly across the traveled distance of 2 microns, which clearly exceeds the wavelength. Panel **(c)** shows the corresponding image of a spin wave excited at 1.46 GHz, panel **(d)** displays

an enlarged image of the center region. In addition to the plane waves originating from the domain wall, spin waves with circular wave fronts can be seen. The latter are emitted from the vortex core (c.f. [18]). Using such images, we obtain the dispersion relation for the various types of waves.

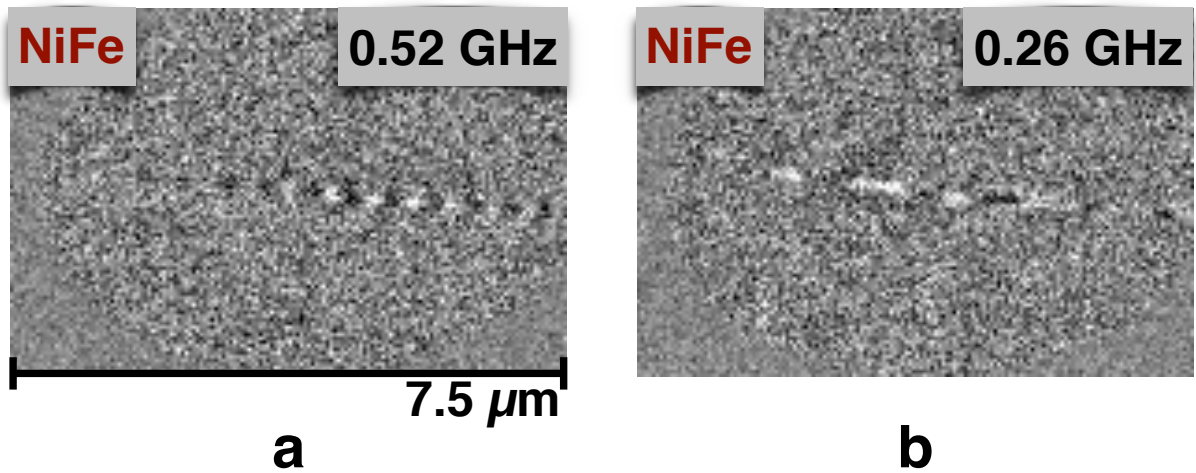

**Figure 4: Spin waves in the domain wall.** At excitation frequencies below a certain threshold, no spin waves are present in the domain regions. However, spin waves confined to the domain wall are observed that originate from the vortex core and travel towards the rim of the disc. Panel **(a)** and **(b)** display such waves (TR-STXM of the NiFe layer magnetisation out-of-plane component) excited at 0.52 and 0.26 GHz, respectively, with accordingly changing wavelengths.

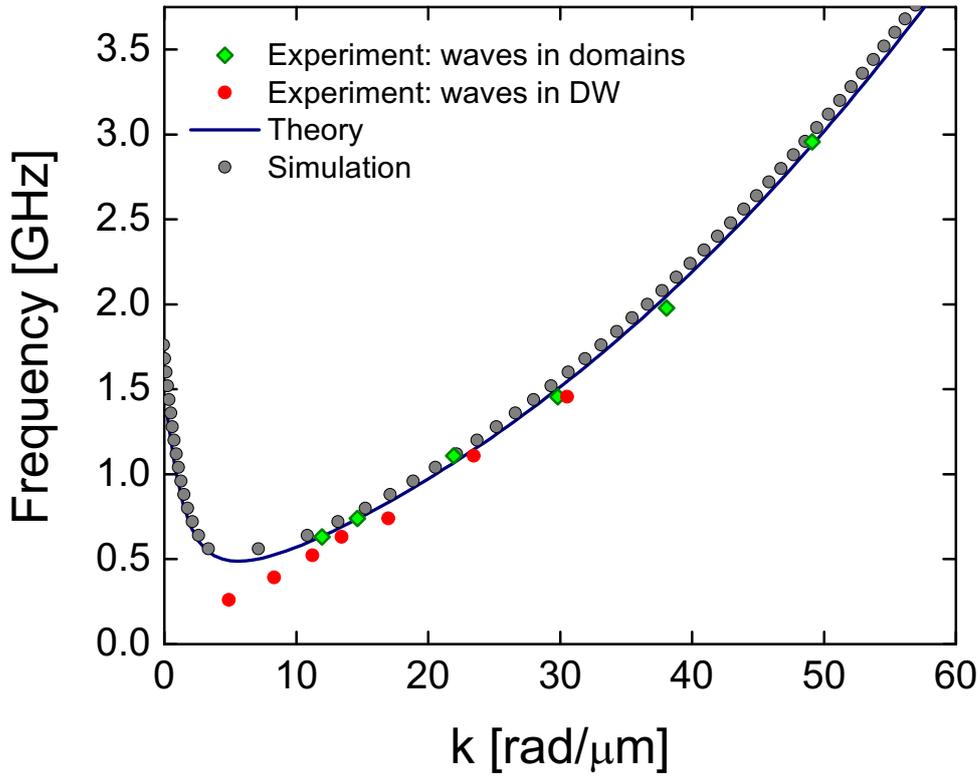

**Figure 5: Spin waves dispersion relations.** From the experiment, we obtain the dispersion relations for spin waves propagating in the domains (full green diamonds) and waves confined to the domain wall (full red dots). In addition, we show the plane wave dispersions calculated using our model (blue continuous line) and micromagnetic simulations (grey dots), which are in good agreement. Assuming an interlayer exchange coupling of -0.1 mJ/m$^2$ and a CoFeB uniaxial anisotropy of 3 kJ/m$^3$, we obtain a reasonable agreement between the numeric results and the measured plane wave dispersion. The key difference of the plane waves and the waves confined to and propagating through the domain walls is the existence of a frequency gap in the plane wave dispersion, while for the wall waves, such a gap is absent, or too small to play a role here. These results explain why it is possible to selectively excite spin waves in the domain wall.

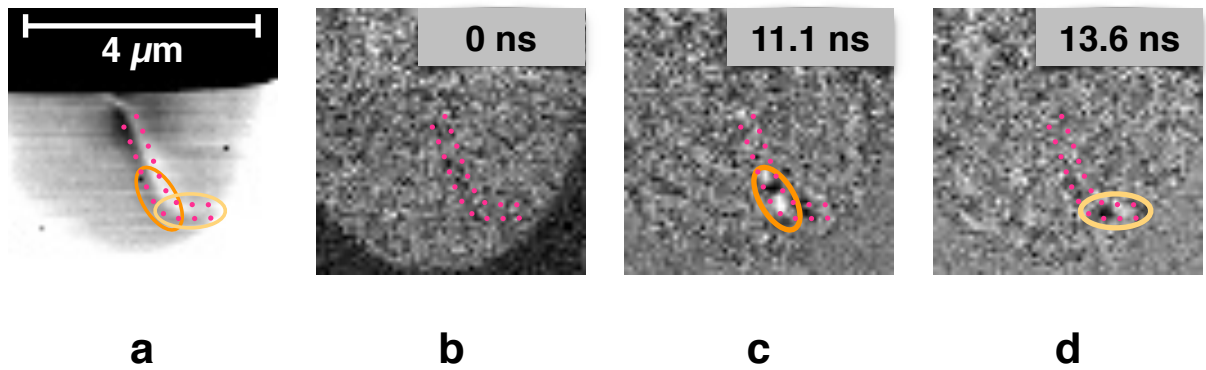

**Figure 6: Domain walls as waveguides.** We demonstrate the possibility of harnessing domain walls as waveguides for magnonic excitations by sending a spin wave packet around a domain wall curving around a corner (TR-STXM images). Panel **(a)** displays an in-plane magnetic contrast image of the domain wall, where the orange and yellow ellipses mark two regions in front and behind the curve, respectively. Panels **(b)** to **(d)** show snapshots of a spin wave packet excited by a magnetic field pulse, which are taken at different instances of time after the pulse. At 11.1 ns after the pulse **(c)**, the wave packet has reached the region in front of the curve, as indicated by the orange ellipse. 2.5 ns later, the wave packet has traveled around the corner **(d)**.

## Methods

**Sample fabrication.**

The samples were prepared on x-ray transparent silicon-nitride membrane substrates with a thickness of 200 nm. Multilayer films of $Ni_{81}Fe_{19}/Ru/Co_{40}Fe_{40}B_{20}/Al$ were deposited by magnetron sputtering onto these, where Al (5 nm) is serving as a capping layer for oxidation protection. The thicknesses of the ferromagnetic layers, NiFe and CoFeB, were determined by transmission electron microscopy to be 44.9 nm and 46.6 nm, respectively. The Ru spacer (0.8 nm nominal) between them mediates an antiferromagnetic interlayer exchange coupling [44] according to the hard axis magnetization reversal of extended multilayer stacks, which was measured by magneto-optic Kerr effect (MOKE) [52]. Additional MOKE measurements on corresponding single layer films revealed collinear uniaxial magnetic anisotropies of 0.2 kJ/m$^3$ for NiFe and 1.1 kJ/m$^3$ for CoFeB. However, in order to reproduce the experimentally found static magnetic configuration in the micromagnetic simulations, a significantly larger value around 3 kJ/m$^3$ is required for CoFeB. This elevated value for the in-plane uniaxial anisotropy in CoFeB can be attributed to strain, which in our case is caused by the contact with the waveguide. In fact, CoFeB is known for its sensitivity to strain and the orientation of the experimentally observed magnetic pattern with respect to the waveguide is consistent with this interpretation [47]. The patterning of the microelements was realized by electron beam lithography (EBL) and consecutive ion beam etching. Upon an initial oxygen plasma treatment for adhesive purposes, a negative resist (MA-N 2910) was spun onto the multilayer films. In a second step, the microelements were exposed by EBL. The samples were then developed for 300 s in MA-D 525 and rinsed in de-ionized water. Finally, the samples were exposed to an argon ion beam at two different angles (85° and 5°) for physically etching the magnetic microelements out of the continuous films. Remaining resist was removed by aceton and a second oxygen plasma treating. For magnetic field excitation, a copper strip of 200 nm thickness was fabricated on top of the microelements by means of EBL, electron beam evaporation deposition, and lift-off processing [18]. The patterned microstrip has a width of 5 µm, hence the resulting magnetic Oersted field from a flowing electric current of one mA can be estimated to $\mu_0 H = 4\pi * 10^{-2}$ mT.

**Time-resolved STXM.**

The magnetic orientation in the multilayer microelement investigated was imaged by means of synchrotron based scanning transmission x-ray microscopy (STXM) [53]. Here, a Fresnel zone plate is used to focus a monochromatic x-ray beam onto the sample. The locally transmitted x-ray intensity is then measured by a single pixel detector, hence raster scanning the sample yields a two-dimensional absorption image with approximately 25 nm lateral resolution. Using furthermore circularly polarized x-rays allows for exploiting x-ray magnetic circular dichroism (XMCD) [54] leading to a magnetic contrast. As XMCD only occurs at the element specific resonant absorption edges, the magnetic signal from both ferromagnetic layers, NiFe and CoFeB, can be separated by tuning the incident x-rays to the corresponding $L_3$ energies, Ni $L_3$ ~853 eV and Co $L_3$ ~778 eV, respectively. On the other hand, a collective signal from both layers can be collected from the Fe $L_3$ edge at ~708 eV since both layers contain Fe. The magnetic contrast acquired is proportional to the projection of the magnetic orientation **m** = **M**/M on the x-ray propagation direction $\mathbf{e}_k$. Therefore, in normal incidence, the STXM setup is sensitive to the perpendicular magnetization component, while an inclined sample mounting also allows for detecting in-plane magnetization components.

The magnetization dynamics of the multilayer microelements was imaged stroboscopically by means of time-resolved STXM. This method utilizes the specific time structure of the incident x-ray pulses, i.e. 2 ns repetition rate at ~100 ps effective pulse length. Each incoming signal (photon or no photon transmitted) is routed after every pulse to a periodic counting register of a field programmable gate array. Here the number of registers ($Q$) sets the maximum non-stroboscopic observation period ($Q*2$ ns), while the number of excitation repetitions in this period ($J$) sets the nominal time resolution as well as the excitation frequency in case of a continuous sinusoidal excitation. The excitation current was measured both in front of and behind the sample by means of -20 dB pick-off tees through an oscilloscope.

**Micromagnetic simulations.**

Micromagnetic simulations based on the time integration of the Landau-Lifshitz-Gilbert [55,56] equation were carried out using the code MuMax$^3$ [57]. The simulations were performed to compute the spin wave dispersion relations in the coupled layer system. The ferromagnetic layers are homogeneously magnetised and the dispersion relations are calculated in a thin film approach. Therefore, the system was discretised into (4096,16,115) (x,y,z) cells and periodic boundary conditions were applied along the y – direction, which corresponds to the direction of equilibrium magnetisation. The thickness of the individual layers and the spacer were chosen according to TEM measurements. This results in a cell size along the z – axis of 0.8 nm. The material parameters used in the micromagnetic simulations are as follows: For NiFe, the respective values of saturation magnetisation, exchange stiffness and uniaxial in-plane anisotropy are: $M_s^{Py} = 800 \text{ kA/m}$, $A_{xc}^{Py} = 7.5 \text{ pJ/m}$ [58] and $K_U^{Py} = 200 \text{ J/m}^3$. For the CoFeB layer, we used $M_s^{CoFeB} = 1250 \text{ kA/m}$, $A_{xc}^{CoFeB} = 12 \text{ pJ/m}$ [59] and $K_U^{CoFeB} = 3000 \text{ J/m}^3$. The interlayer exchange coupling is $J = -0.1 \text{ mJ/m}^2$. The Gilbert damping constant α for CoFeB and NiFe is chosen to 0.008 and 0.01, respectively. The prevent reflection of spin waves from the edges, the damping was increased linearly to 0.065 for both layers.

An out-of-plane sinusoidal excitation field with a fixed frequency was applied in a 100 nm wide region in the center of the system. After the system reached the dynamic equilibrium, the magnetisation configuration was stored. To extract the wave number for each frequency a spatial fast-Fourier transform along the x – direction of the system was performed. The corresponding dispersion relations are shown in Fig. 5 as grey full dots and are in good agreement with the result from the model calculations. Additional simulations were performed in order to compare the effects of oscillatory magnetic fields applied in- and out-of-plane, in each case perpendicular to the magnetization. The simulations clearly show that the bilayer system is more susceptible to out-of-plane field perturbations. This result can be understood taking into account the fact that the excited collective mode exhibits an in-phase oscillation of the perpendicular, yet anti-phase oscillation of the in-plane magnetisation component, and thus couples more efficiently to driving fields oriented perpendicular to the sample plane. There exists another type of collective mode in the system, which exhibits an in-phase oscillation of the in-plane component, however this mode resides at higher frequency values than the measured ones.

Due to the absence of domain walls, the thin film approach above cannot reproduce the observed one-dimensional spin-wave dispersion within the walls and the emission of planar waves from them. To gain insight into these phenomena an elliptical bilayer with a short and long axis of 2.16 µm and 3.24 µm, respectively, was simulated at a discretisation of 648 x 432 x 115 cells. The material parameters were chosen to be the same as for the thin film approach, but for the slightly modified constants $A_{xc}^{Py} = 10.5$ pJ/m, $A_{xc}^{CoFeB} = 13$ pJ/m, $K_u^{CoFeB} = 5$ kJ/m$^3$ and $J_{IEC} = -0.3$ mJ/m$^2$. To simulate the excitation of spin-wave dynamics a spatially homogeneous sinc(*t*)-pulse with a cut-off frequency of 10 GHz was utilised. To obtain the dynamic response the magnetisation data was Fourier-transformed and filtered afterwards to extract the desired frequency.

**Data availability.**

The datasets generated and/or analysed during the current study are available from the corresponding authors on reasonable request.

# References

[1] Bloch, F. Zur Theorie des Ferromagnetismus. *Z. Phys.* **61** 206 (1930).

[2] Holstein, T. & Primakoff, H. Field Dependence of the Intrinsic Domain Magnetization of a Ferromagnet. *Phys. Rev.* **58** 1098 (1940).

[3] Dyson, FJ. General Theory of Spin Wave Interactions. *Phys. Rev.* **102** 1217 (1956).

[4] Kruglyak, V. V., Demokritov, S. O. & Grundler, D. Magnonics. *J. Phys. D: Appl. Phys.* **43** 264001 (2010).

[5] Chumak, A. V., Serga, A. A. & Hillebrands, B. Magnon transistor for all-magnon data processing. *Nat. Commun.* **5** 4700 (2014).

[6] Chumak, A.V., Vasyuchka, V.I., Serga, A.A., & Hillebrands, B. Magnon Spintronics. *Nat. Phys.* **11** 453 (2015).

[7] Lan, J., Yu, W., Wu, R. & Xiao, J. Spin-Wave Diode. *Phys. Rev. X* **5** 041049 (2015).

[8] Gurevich, A.G. & Melkov, G.A. Magnetization Oscillations and Waves. *New York: CRC* (1996).

[9] Acoustic Surface Waves, Ed. A.A. Oliner, Springer-Verlag Berlin Heidelberg 1978

[10] D.R.Morgan, Surface acoustic wave devices and applications: 1. Introductory review, Ultrasonics **11**, 121-131, (1973).

[11] Lee, K.-S., Choi, S. & Kim, S.-K. Radiation of spin waves from magnetic vortex cores by their dynamic motion and annihilation processes. *Appl. Phys. Lett.* **87** 192502 (2005).

[12] Demokritov, S. O., Demidov, V. W., Dzyapko, O., Melkov, G. A., Serga, A. A., Hillebrands, B. & Slavin, A. N. Bose-Einstein condensation of quasi-equilibirum magnons at room tempertature under pumping. *Nature* **443** 430 (2006).

[13] Demidov, V. E., Urazhdin, S. & Demokritov, S. O. Direct observation and mapping of spin waves emitted by spin-torque nano-oscillators. *Nature Mater.* **9** 984 (2010).

[14] Madami, M., Bonetti, S., Consolo, G., Tacchi, S., Carlotti, G., Gubbiotti, G., Mancoff, F. B., Yar, M. A. & Åkerman, J. Direct observation of a propagating spin wave induced by spin-transfer torque. *Nat. Nanotech.* **6** 635 (2011).

[15] Yu, H., Duerr, G., Huber, R., Bahr, M., Schwarze, T., Brandl, F. & Grundler, D. Omnidirectional spin-wave nanograting coupler. *Nat. Commun.* **4** 2702 (2013).

[16] Urazhdin, S., Demidov, V. E., Ulrichs, H., Kendziorczyk, T., Kuhn, T., Lethold, J., Wilde, G. & Demokritov, S. O. Nanomagnetic devices based on the spin-transfer torque. *Nat. Nanotech.* **9** 509 (2014).

[17] Yu, H., d' Allivy Kelly, O., Cros, V., Bernard, R., Bortolotti, P., Anane, A., Brandl, F., Heimbach, F. & Grundler, D. Approaching soft X-ray wavelengths in nanomagnet-based microwave technology. *Nat. Commun.* **7** 11255 (2016).

[18] Wintz, S., Tyberkevych, V., Weigand, M., Raabe, J., Lindner, J., Erbe, A., Slavin, A.N. & Fassbender, J. Magnetic vortex cores as tunable spin wave emitters. Nat. Nanotech. **11** 948 (2016).

[19] Davies, C. S., Poimanov, V. D. & Kruglyak V. V. Mapping the magnonic landscape in patterned magnetic structures. *Phys. Rev.* **96** 094439 (2017).

[20] Mozooni, B. & McCord, J. Direct observation of closure domain wall mediated spin waves. *Appl. Phys. Lett.* **107** 042402 (2015).

[21] Hämäläinen, S. J., Brandl, F., Franke, K. J. A., Grundler, D. & van Dijken, S. Tunable Short-Wavelength Spin-Wave Emission and Confinement in Anisotropy-Modulated Multiferroic Heterostructures. *Phys. Rev. Appl.* **8** 014020 (2017).

[22] Wiele, B.V., Hämäläinen, S. J., Baláz, P., Montoncello, F. & Dijken, S.V. Tunable short-wavelength spin wave excitation from pinned magnetic domain walls. *Sci. Rep.* **6** 21330 (2016).

[23] Whitehead, N.J., Horsley, S.A.R., Philbin, T.G., Kuchko, A.N. & Kruglyak, V.V. Theory of linear spin wave emission from a Bloch domain wall. *Phys. Rev. B* **96** 064415 (2017).

[24] Voto, M., Lopez-Diaz, L. & Martinez, E. Pinned domain wall oscillator as tunable direct current spin wave emitter. *Sci. Rep.* **7** 13559 (2017).

[25] Hermsdoerfer, S. J., Schultheiss, H., Rausch, C., Schäfer, S., Leven, B., Kim, S.-K. & Hillebrands, B. A spin-wave frequency doubler by domain wall oscillation. *Appl. Phys. Lett.* **94** 223510 (2009).

[26] Bonetti, S., Kukreja, R., Chen, Z., Macià, F. Hernàndez, J. M., Eklund, A., Backes, D., Frisch, J., Katine, J., Malm, G., Urazhdin, S., Kent, A. D., Stöhr, J., Ohldag, H. & Dürr, H. A. Direct observation and imaging of a spin-wave soliton with *p*-like symmetry. *Nat. Commun.* **6** 8889 (2015).

[27] Dieterle, G., Förster, J., Stoll, H., Semisalova, A. S., Finizio, S., Gangwar, A., Weigand, M., Noske, M., Fähnle, M., Bykova, I., Gräfe, J., Bozhko, D. A., Musiienko-Shmarova, H. Yu., Tiberkevich, V., Slavin, A. N., Back, C. H., Raabe, J., Schütz, G. & Wintz, S. Coherent excitation of heterosymmetric spin waves with ultrashort wavelengths. arXiv:1712.00681 (2017).

[28] Holländer, R. B., Müller, C., Schmalz, J., Gerken, M. & McCord, J. Magnetic domain walls as broadband spin wave and elastic magnetisation wave emitters. *Sci. Rep.* **8** 13871 (2018).

[29] Sluka, V., Weigand, M., Kakay, A., Erbe, A., Tyberkevych, V., Slavin, A., Deac, A., Lindner, J., Fassbender, J., Raabe, J. & Wintz, S. Stacked topological spin textures as emitters for multidimensional spin wave modes. *IEEE International Magnetics Conference (INTERMAG), Beijing, 11-15 May, 2015* (IEEE, New York, 2015)

[30] Garcia-Sanchez, F., Borys, P., Soucaille, R., Adam, J.-P., Stamps, R.L. & Kim, J.-V. Narrow Magnonic Waveguides Based on Domain Walls. *Phys. Rev. Lett.* **114** 247206 (2015).

[31] Wagner, K., Kákay, A., Schultheiss, K., Henschke, A., Sebastian, T. & Schultheiss, H. Magnetic domain walls as reconfigurable spin-wave nanochannels. *Nat. Nanotechnol.* **11** 432 (2016).

[32] Xing, X. & Zhou, Y. Fiber optics for spin waves. *NPG Asia Materials* **8** e246 (2016).

[33] Winter, J.M. Bloch Wall Excitation. Application to Nuclear Resonance in a Bloch Wall. *Phys. Rev.* **124** 452 (1961).

[34] Albisetti, E., Petti, D., Sala, G., Silvani, R., Tacchi, S., Finizio, S., Wintz, S., Caló, A., Zheng, X., Raabe, J., Riedo, E. & Bertacco R. Nanoscale spin-wave circuits based on engineered reconfigurable spin-textures. Comms. Phys. **1** 56 (2018).

[35] Aliev, F. G., Awad, A. A., Dieleman, D., Lara, A., Metlushko, V. & Guslienko, K. Y. Localized domain-wall excitations in patterned magnetic dots probed by broadband ferromagnetic resonance. Phys. Rev. B **84** 144406 (2011).

[36] Stoll, H., Puzic, A., van Wayenberge, B., Fischer, P., Raabe, J., Buess, M., Haug, T., Höllinger, R., Back, C., Weiss, D. & Denbeaux, G. High-resolution imaging of fast magnetization dynamics in matnetic nanostructures. Appl. Phys. Lett. **84** 3328 (2004).

[37] Vogt, K., Fradin, F. Y., Pearson, J. E., Sebastian, T., Bader, S. D., Hillebrands, B., Hoffmann, A. & Schultheiss, H. Realization of a spin-wave multiplexer. *Nat. Commun.* **5** 3727 (2014).

[38] Gruszecki, P., Kasprzak, M., Serebryannikov, A.E., Krawczyk, M. & Smigaj, W. Microwave excitation of spin wave beams in thin ferromagnetic films. *Sci. Rep.* **6** 22367 (2016).

[39] Albisetti, E., Petti, D., Pancaldi, M., Madami, M., Tacchi, S., Curtis, J., King, W. P., Papp, A., Csaba, G., Porod, W., Vavassori, P., Riedo, E., Bertacco, R. Nanopatterning reconfigurable magnetic landscapes via thermally assisted scanning probe lithography. *Nat. Nanotech.* **11** 545 (2016).

[40] Haldar, A., Kumar, D. & Adeyeye, A.O. A reconfigurable waveguide for energy-efficient transmission and local manipulation of information in a nanomagnetic device. *Nat. Nanotech.* **11** 437 (2016).

[41] Shinjo, T., Okuno, T., Hassdorf, R., Shigeto, K. & Ono, T. Magnetic Vortex Core Observation in Circular Dots of Permalloy. *Science* **289** 930 (2000).

[42] Thiele A. A. Steady-State of Motion of Magnetic Domains. *Phys. Rev. Lett.* **30** 230 (1973).

[43] Choe, S.-B., Acremann, Y., Scholl, A., Bauer, A., Doran, A., Stöhr, J. & Padmore, H. A. Vortex core-driven magnetization dynamics. *Science* **304** 420 (2004).

[44] Grünberg, P., Schreiber, R., Pang, Y., Brodsky, M. B. & Sowers, H. Layered Magnetic Structures: Evidence for Antiferromagnetic Coupling of Fe Layers across Cr Interlayers. *Phys. Rev. Lett.* **57** 2442 (1986).

[45] Labrune, M. & Miltat, J. Wall structures in ferro/antiferromagnetic exchange-coupled bilayers: a numerical micromagnetic approach. *J. Magn. Magn. Mater.* **151** 231 (1995).

[46] Stamps, R. L., Carriço, A. S. & Wigen, P. E. Domain-wall resonance in exchange-coupled magnetic films. *Phys. Rev. B* **55** 6473 (1997).

[47] Wang, D., Nordman, C., Qian, Z., Daughton, J. M. & Myers, J. Magnetostriction effect of amorphous CoFeB thin films and application in spin-dependent tunnel junctions. *J. Appl. Phys.* **97** 10C906 (2005).

[48] Grünberg, P. Magnetostatic spinwave modes of a heterogeneous ferromagnetic double layer. *J. Appl. Phys.* **52** 6824 (1981).

[49] Damon, R. W. & Eshbach, J. R. Magnetostatic modes of a ferromagnetic slab. *J. Phys. Chem. Solids* **19** 308 (1961).

[50] Cho, J., Kim, N.-H., Lee, S., Kim, J.-S., Lavrijsen, R., Solignac, A., Yin, Y., Han, D.-S., van Hoff, N. J. J., Swagten, H. J. M., Koopmans, B & You, C.-Y. Thickness dependence of the interfacial Dzyaloshinskii-Moriya interaction in inversion symmetry broken systems. Nat. Commun. **6** 7635 (2015).

[51] Goldstone, J., Salam, A., & Weinberg S. Broken Symmetries. Phys. Rev. **127** 965 (1962).

[52] Wintz, S., Strache, T., Körner, M., Bunce, C., Banholzer, A., Mönch, I., Mattheis, R., Raabe, J., Quitmann, C., McCord, J., Erbe, A., Lenz, K. & Fassbender, J. Control of vortex pair

states by post-deposition interlayer exchange coupling modification. *Phys. Rev. B* **85** 134417 (2012).

[53] Raabe, J., Tzetkov, G., Flechsig, U., Böge, M., Jaggi, A., Sarafimov, B., Vernooij, M. G. C., Huthwelker, T., Ade, H., Kilcoyne, D., Tyliszczak, T., Fink, R. H. & Quitmann, C. PolLux: A new facility for soft x-ray spectromicroscopy at the Swiss Light Source. Rev. Sci. Instrum. **79** 113704 (2008).

[54] Schütz, G., Wagner, W., Wilhelm, W., Kienle, P., Zeller, R., Frahm R. & Materlik, G. Absorption of circularly polarized x rays in iron. *Phys. Rev. Lett.* **58** 737 (1987).

[55] Landau, L. & Lifshits E. On the theory of the dispersion of magnetic permeability in ferromgnetic bodies. *Phys. Zeitsch. der Sow.* **8** 135 (1935).

[56] Gilbert, T. L. A Lagrangian formulation of the gyromagnetic equation of the magnetization Field. *Phys. Rev.* **100** 1243 (1955).

[57] Vansteenkiste, A., Leliaert, J., Dvornik, M., Helsen, M., Garcia-Sanchez, F. & Van Waeyenberge, B. The design and verification of MuMax3. *AIP Advances* **4** 107133 (2014).

[58] Wei, J., Zhu, Z., Song, C., Feng, H., Jing, P., Wang, X., Liu, Q. & Wang, J. Annealing influence on the exchange stiffness constant of Permalloy films with stripe domains. *J. Phys. D: Appl. Phys.* **49** 265002 (2016).

[59] Conca, A., Papaioannou, E.Th., Klingler, S., Greser, J., Sebastian, T., Leven, B., Lösch, J. & Hillebrands, B. Annealing influence on the Gilbert damping parameter and the exchange constant of CoFeB films. *Appl. Phys. Lett.* **104** 182407 (2014).

## Acknowledgements

We would like to thank B. Sarafimov, B. Watts, and M. Bechtel for experimental support at the STXM beamlines as well as C. Fowley, K. Kirsch, B. Scheumann, and C. Neisser for their help in sample fabrication. Most of the experiments were performed at the Maxymus endstation at BESSY2, HZB, Berlin, Germany. We thank HZB for the allocation of synchrotron radiation beamtime. Some experiments were performed at the PolLux endstation at SLS, PSI, Villigen, Switzerland. Pollux is financed by BMBF via contracts 05KS4WE1/6 and 05KS7WE1. Support by the Nanofabrication Facilities Rossendorf at IBC, HZDR, Dresden, Germany is gratefully acknowledged. V.S. and A.D. acknowledge funding from the Helmholtz Young Investigator Initiative under Grant VH-N6-1048. R.A.G. acknowledges financial support from FONDECYT Iniciacion 11170736 and 1161403. A.R.M. acknowledges funding from FONDECYT 3170647; funding from Basal Program for Centers of Excellence, Grant FB0807 CEDENNA, CONICYT is also acknowledged. V.T. and A.S. acknowledge support from the National Science Foundation of the USA under the Grants No.EFMA-1641989 and No. ECCS-1708982 and from the DARPA M3IC grant under the Contract No.W911-17-C-0031. S.W. acknowledges funding from the European Community's Seventh Framework Programme (FP7/2007-2013) under grant agreement n.°290605 (PSI-FELLOW/COFUND).

## Author contributions

V.S., M.W., and S.W. performed the STXM measurements. T.S., T.W., A.K., and S.W. did the micromagnetic simulations. R.A.G, A.R.M, and P.L calculated the spin wave dispersion relation. R.M. and S.W. supervised the sample preparation. V.S. and S.W. wrote the manuscript. All authors contributed to the discussion of the results and commented on the manuscript.

# Supplementary Information

# Emission and Propagation of Multi-Dimensional Spin Waves with nanoscale wavelengths in Anisotropic Spin Textures

V. Sluka[1,a,*], T. Schneider[1], R. A. Gallardo[2,3], A. Kakay[1], M. Weigand[4], T. Warnatz[1,b], R. Mattheis[5], A. Roldan-Molina[6], P. Landeros[2,3], V. Tiberkevich[7], A. Slavin[7], G. Schütz[4], A. Erbe[1], A. Deac[1], J. Lindner[1], J. Raabe[8], J. Fassbender[1,9], and S. Wintz[1,8,*]

[1] Helmholtz-Zentrum Dresden-Rossendorf, 01328 Dresden, Germany
[2] Universidad Técnica Federico Santa María, 2390123 Valparaíso, Chile
[3] Center for the Development of Nanoscience and Nanotechnology (CEDENNA), 917-0124 Santiago, Chile
[4] Max-Planck-Institut für Intelligente Systeme, 70569 Stuttgart, Germany
[5] Leibniz Institut für Photonische Technologien, 07745 Jena, Germany
[6] Universidad de Aysén, Calle Obispo Vielmo 62, Coyhaique, Chile
[7] Oakland University, Rochester, MI 48309, USA
[8] Paul Scherrer Institut, 5232 Villigen PSI, Switzerland
[9] Technische Universität Dresden, 01069 Dresden, Germany
[a] Present address: New York University, New York, NY 10003, USA
[b] Present address: Uppsala Universitet, 75120 Uppsala, Sweden
[*] e-mail: vs1568@nyu.edu; s.wintz@hzdr.de

## (1) Supplementary Movies

**Time-resolved scanning transmission x-ray microscopy**

Selected time-resolved scanning transmission X-ray microscopy measurements are shown as movies (**M1-M6**) with perpendicular magnetic sensitivity for both different samples (#1 and #2) and excitation schemes (continuous wave and pulsed). These movies display the absolute ($\sim m_z$) and normalised ($\sim \Delta m_z$) magnetic contrast at different absorption edges (Fe, Co, Ni $L_3$) as described below:

**M1_s1_Ni_8um-5um_cw1.11GHz_absolute-normalised**
- sample: #1, photon energy: Ni $L_3$ edge, scan size: $8 \times 5 \mu m^2$, scan step: 50 nm,
- excitation: continuous wave N=23, M=51, frequency: 1.11 GHz, time step: 39 ps
- absolute contrast (left), normalised contrast (right)

**M2_s1_Ni_7.5um-x-5um_cw1.46GHz_absolute-normalised**
- sample: #1, photon energy: Ni $L_3$ edge, scan size: $7.5 \times 5 \mu m^2$, scan step: 50 nm,
- excitation: continuous wave N=23, M=67, frequency: 1.46 GHz, time step: 30 ps
- absolute contrast (left), normalised contrast (right)

**M3_ s1_2um-x-2um_cw1.46GHz_absolute-normalised_Ni-Co-Fe**
- sample: #1, scan size: $2 \times 2 \mu m^2$, scan step: 25 nm,
- excitation: continuous wave N=23, M=67, frequency: 1.46 GHz, time step: 30 ps
- absolute contrast (left), normalised contrast (right)
photon energy: Ni $L_3$ edge (top), Co $L_3$ edge (middle), Fe $L_3$ edge (bottom)

**M4_s1_Ni_7.5um-x-5um_absolute-normalised_cw0.52GHz-cw0.26GHz**
- sample: #1, photon energy: Ni $L_3$ edge, scan size: $7.5 \times 5 \mu m^2$, scan step: 50 nm,
- absolute contrast (left), normalised contrast (right)
- excitation: continuous wave N=23, M=24, frequency: 0.52 GHz, time step: 83 ps (top)
- excitation: continuous wave N=23, M=12, frequency: 0.26 GHz, time step: 167 ps (bottom)

**M5_s1_Ni_9um-x-1.5um_pulse0.6ns_50.5ns_absolute-normalised**
- sample: #1, photon energy: Ni $L_3$ edge, scan size: $9 \times 1.5 \mu m^2$, scan step: 50 nm,
- excitation: pulsed N=101, M=4, duration: 600 ps, time step: 500 ps, period: 50.5 ns

- absolute contrast (left), normalised contrast (right)

**M6_s2_Fe_4.5um-x-4.5um_pulse0.6ns_25.1ns_normalised**
- sample: #2, photon energy: Ni $L_3$ edge, scan size: 9x1.5µm$^2$, scan step: 50 nm,
- excitation: pulsed N=101, M=4, duration: 600 ps, time step: 500 ps, period: 50.5 ns
- normalised contrast

**Micromagnetic Simulations**

**M7_sim_ellipse-3.24um-x-2.16um_CoFeB_sinc_FFT404MHz_normalised**
- simulation according to methods section, showing the normalised perpendicular magnetic response (~$\Delta m_z$) of the CoFeB layer at $f$ = 404 MHz

## (2) Spin-wave modes in antiparallel magnetic bilayers

The bilayer system with antiparallel magnetic layers considered in the present work can host spin-wave excitations of acoustic and optical character. This applies both to spin waves in the bulk domains and to waves in the domain wall. The acoustic mode spin waves in the domain wall can be considered as the bilayer-analogue of the modes proposed by *Winter* [33]. The magnetization precession associated with that mode for finite $k$ results in a string-like in-phase oscillation of both individual domain walls. In the limit $k \to 0$ this mode corresponds to the coherent domain wall oscillations as described by Stamps *et al*. [46]. As discussed in the main text, due to symmetry, this acoustic domain wall mode has a (near) gap-less dispersion relation. However, the optical domain wall mode, as well as both optical and acoustic plane spin wave modes in the domains exhibit gaps in their dispersion relations.

In the figure below, we show the respective dispersion relations for the optical and acoustic spin waves in the domains, comparing results obtained from theory and from micromagnetic simulations [panel (a)]. The pink symbols represent the two lowest $k$-vector data points from the experiment. The optical and acoustic modes can be easily compared when imagining looking at the individual layers' magnetizations from the perspective such that the magnetization points directly at the viewer. In that case, each layer's magnetization precesses with the same sense of rotation, so that the terms "optical" and "acoustic" refer to the different phase relationships (180 and zero degrees, respectively). In addition to the phase shift, the waves of a particular mode differ also with respect to the propagation direction, as can be seen from the asymmetry with respect to the sign of the $k$ coordinate. The resulting four combinations of phase relationships and sign of $k$ are illustrated by the inset schematics. Panel [b] displays the respective mode profiles along the coordinate perpendicular to the sample plane. Note that a completely strict separation between optical and acoustic mode is only possible in a system with two identical layers. Yet, for the layers and frequencies considered here, phase deviations from the pure coupled optic and acoustic mode schemes are very small and therefore neglected. As can be seen in the figure, the optical mode resides at much much higher frequencies than the acoustic mode. At the same time, both modes (in particular the accoustic one) are strongly non-reciprocal, in the sense that waves propagating into opposite directions have different wavelengths at the same frequency. Therefore the two branches ($\pm k$) can be defined as slow branch and fast branch with respect to their group velocity ($d\omega/dk$). Note that there is a frequency gap (accoustic: ~500 MHz *vs.* optical: ~3 GHz) for both modes, and that the frequency minimum of the accoustic mode is shifted to finite $k$ to the slow branch side.

In our experiment, however, we only observe waves belonging to the slow branch of the accoustic mode. One reason for this is that the optical mode only exists at frequencies above those addressed in our experiment. Secondly, for the accoustic mode, the fast branch is not observed as the expected amplitudes are relatively low and because the wavelengths at the experimental frequencies are of the order of or beyond the sample size itself.

For the one-dimensional waves confined to the domain walls, the situation is similar. The accoustic and optical spin-wave modes are also non-reciprocal. As mentioned above, the acoustic mode is gapless due to the Goldstone theorem and in agreement with earlier predictions, while the domain wall optical mode still exhibits a frequency gap. Also for the spin waves confined to the domain walls, we only abserve the slow branch of the accoustic mode.

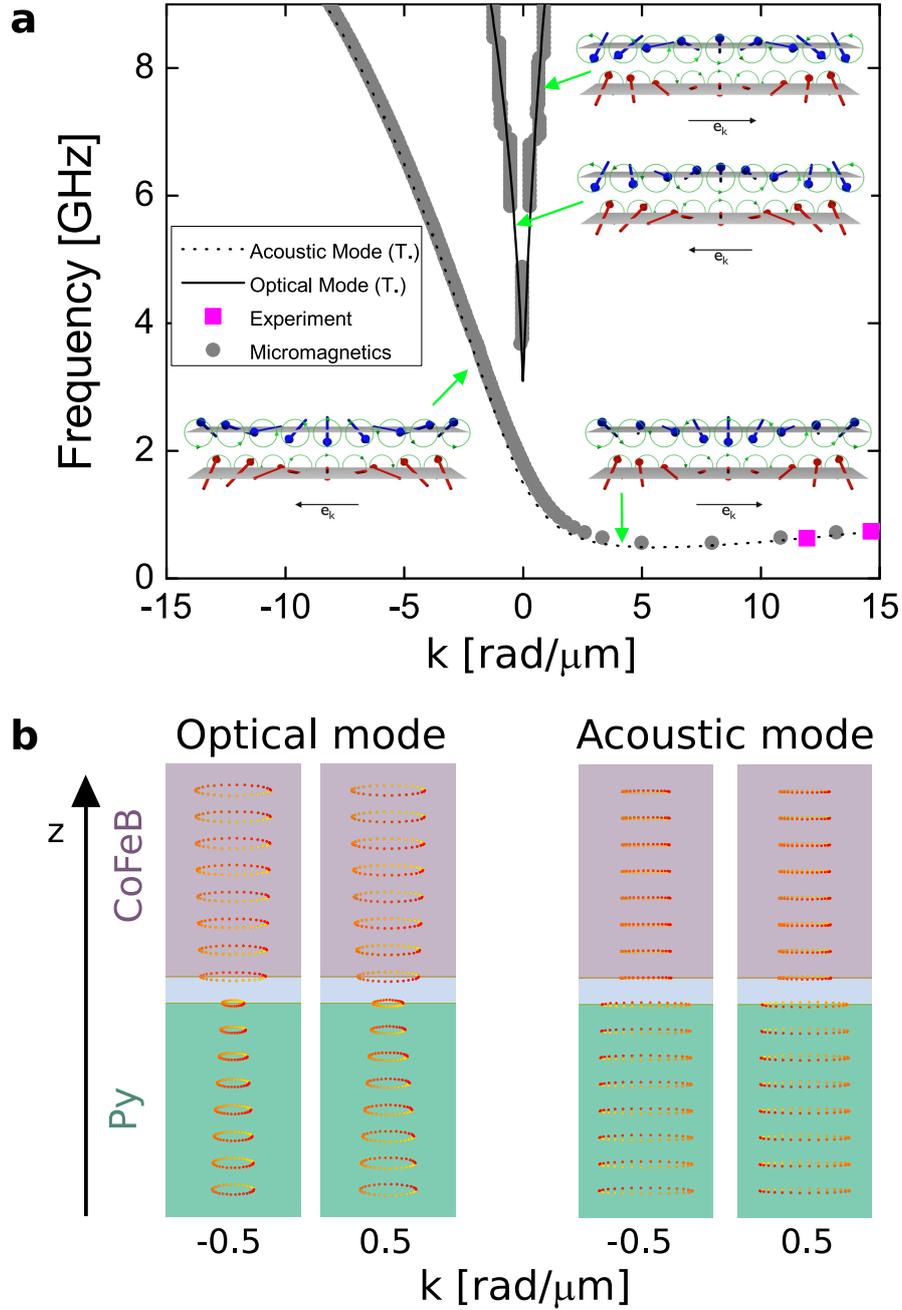

**Figure S1: Dispersion and mode profiles for plane waves in the domains.** Panel (a) shows the dispersion relations for acoustic and optical modes in the domains. Both modes exhibit a frequency gap, and the point of lowest frequency of the acoustic mode is shifted towards positive $k$. The modes are also nonreciprocal, as their dispersion relations are not symmetrical with respect to $k = 0$. The insets schematically depict the phase relationship and propagation direction for the respective mode and sign of $k$. Panel (b) shows the mode profiles along the perpendicular to plane $z$-coordinate, for optical and acoustic modes and opposite $k$-polarities. The colored dots represent the tip of the dynamic component of the local magnetization vector. The color scheme of the dots varying from yellow to red refers to the phase.

## (3) Spin-wave excitation efficiency

In order to compare the efficiency of spin wave excitation by spin textures to that of patterned antennas (such as co-planar wave guides), we assume that an antenna of rectangular cross-section and infinite length lies on top of the magnetic layer system along the x-direction. The layer magnetizations are collinear with the x-direction, and z denotes the coordinate in the out-of-plane direction. Then the Oersted field emanating from the antenna has the form $\boldsymbol{H}(y,z)\cos(\omega t)$. We further assume that it is the Fourier-component $\boldsymbol{H}(k_y, z)\cos(\omega t)$ that is exciting the spin wave with wave vector $\boldsymbol{k} = k_y \boldsymbol{e}_y$ ($\boldsymbol{e}_y$ denoting the unit vector pointing into the y-direction). As the z-component behaves qualitatively similar, below we focus only on the y-component of that field. It is proportional to the ac current density $j$ sent through the antenna and further determined by the width and height of the wave guide. In Figure S2, we display $H_y(k_y, z_0)$ for two cases: The first case corresponds to the situation in our experiment (blue lines); the antenna width and height are 5 μm and 200 nm, respectively. In the second case (orange), the antenna width is reduced to 200 nm. The value of $z_0$ is chosen to 46 nm below the antenna in each case, which approximately corresponds to the centre of the magnetic layer stack. The current densities are equal in each case.

For the first case, it can be seen that $H_y(k_y, z_0)$ falls off drastically from low to high k-vectors, with an oscillatory k-dependence superimposed, so that at wave lengths such as observed in the experiment ($k_y > 10$ rad/μm and higher), the excitation amplitude is only a fraction of its maximum value (that of the spatially homogeneous component). In the second case of the narrower antenna, the fall-off towards higher k is less dramatic, but the overall amplitude is so low that around $k_y = 10$ rad/μm, the two cases are actually comparable (c.f. the inset Figure S2). The excitation by spin textures as in our experiment is more efficient because it uses the very strong spatially inhomogeneous internal fields arising from dynamics of the spin textures (driven by the homogeneous external field), which in turn excite the high-$\boldsymbol{k}$ spin waves.

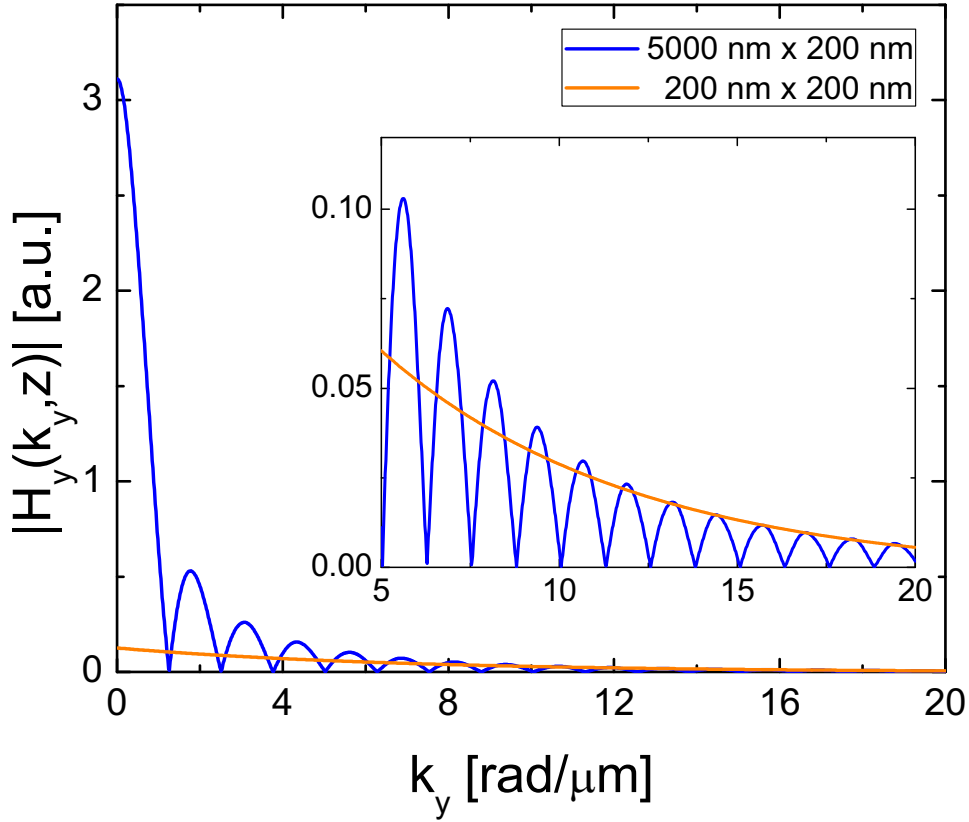

**Figure S2: Spatial Fourier transform of the in-plane magnetic field component emanating from an antenna.** Two cases are compared: The first is similar to the experiment, with the antenna wire exhibiting a width of 5 μm and a height of 200 nm (blue). In the second case (orange), width and height are both equal to 200 nm. The inset compares the field amplitudes at wave vectors similar to the ones of the spin waves in the experiment (> 10 rad/μm). At these **k**, the field amplitudes of broad and narrow antenna are actually comparable, but only correspond to a fraction of the homogeneous field component generated by the wide antenna. That strong component, however, can be exploited in our experiment to indirectly excite spin waves via the domain walls and vortex cores.

# (4) Analytic calculation of the spin-wave dispersion in coupled bilayers

A schematic diagram of the bilayer structure is shown in Fig. 1. The system is composed of two exchange and dipolarly coupled ferromagnetic (FM) layers (1) and (2) with in-plane homogenous magnetization $\mathbf{M}^{(1)}$ and $\mathbf{M}^{(2)}$, respectively. The layers have different magnetic parameters and thicknesses, and $s$ denotes the separation between them. A local coordinates system $(X_\nu,Y_\nu,Z_\nu)$ is defined for each layer ($\nu = 1,2$) [see Fig. 1(b)], where the uniform $\mathbf{M}^{(\nu)}$ is pointing along $X_\nu$. Spin waves (SWs) are assumed to propagate along the $x$-axis. Thus, for the wave-vector $\mathbf{k} = k\hat{x}$, a positive (negative) $k$ represents SW propagation in the positive (negative) $x$ direction. The angle $\varphi_\nu$ corresponds to the equilibrium angle of layer ($\nu$) measured from $\hat{x}$, while $\varphi_\mathrm{h}$ is the field angle.

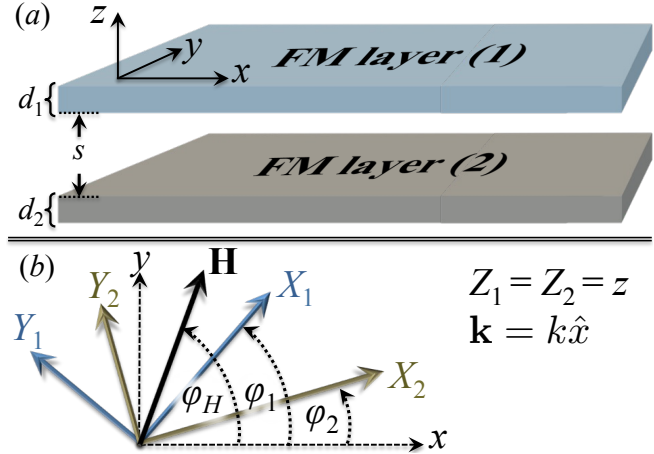

FIG. 1. (a) Overview of the bilayer system. Coordinates ($x$,$y$,$z$) are fixed and both FM layers are separated by a distance $s$. In (b), local coordinates $(X_\nu,Y_\nu,Z_\nu)$ are defined according to the equilibrium magnetization of layer ($\nu$).

The temporal evolution of the system is described by the Landau-Lifshitz (LL) equation

$$\dot{\mathbf{M}}^{(\nu)}(\mathbf{r},t) = -\gamma \mathbf{M}^{(\nu)}(\mathbf{r},t) \times \mathbf{H}^{\mathrm{e}(\nu)}(\mathbf{r},t), \tag{1}$$

where the dot denotes time derivative, $\gamma$ is the absolute value of the gyromagnetic ratio, and $\mathbf{H}^{\mathrm{e}(\nu)}(\mathbf{r},t)$ is the effective field acting on layer $\nu$. In the linear response regime, both magnetization and effective field can be written as $\mathbf{M}^{(\nu)}(\mathbf{r},t) = M_\mathrm{s}^{(\nu)}\hat{X}_\nu + \mathbf{m}(\mathbf{r},t)$ and $\mathbf{H}^{\mathrm{e}(\nu)}(\mathbf{r},t) = H_{X_\nu}^{\mathrm{e}0} + \mathbf{h}^\mathrm{e}(\mathbf{r},t)$ (see Fig. 1). Thus, the linearized equations of motion (1) become

$$i\frac{\omega}{\gamma}m_{Y_\nu}(\mathbf{r}) = -m_{Z_\nu}(\mathbf{r})H^{e0}_{X_\nu} + M_s^{(\nu)}h^e_{Z_\nu}(\mathbf{r}), \tag{2a}$$

$$i\frac{\omega}{\gamma}m_{Z_\nu}(\mathbf{r}) = m_{Y_\nu}(\mathbf{r})H^{e0}_{X_\nu} - M_s^{(\nu)}h^e_{Y_\nu}(\mathbf{r}), \tag{2b}$$

where an harmonic dynamic response has been assumed $\mathbf{m}(\mathbf{r},t) = \mathbf{m}(\mathbf{r})e^{i\omega t}$. Also, according to the equilibrium condition $H^{e0}_{Y_\nu} = H^{e0}_{Z_\nu} = 0$. On the other side, for monochromatic spin waves propagating along $x$, the dynamic magnetization components of both FM layers can be written as $\mathbf{m}(x) = \mathbf{m}(k)e^{ikx}$, where $k$ is the wave vector. In the same way, the dynamic effective field is given by $\mathbf{h}^e(x) = \mathbf{h}^e(k)e^{ikx}$. Therefore, the terms $M_s^{(\nu)}h^e_{Y_\nu}(\mathbf{r})$ and $M_s^{(\nu)}h^e_{Z_\nu}(\mathbf{r})$ in Eq. (2) can be expressed as

$$M_s^{(1)}h^e_{Y_1}(x) = [m_{Y_1}(k)T_{Y_1} + m_{Z_1}(k)T_{Z_1} + m_{Y_2}(k)T_{Y_2} + m_{Z_2}(k)T_{Z_2}]e^{ikx},$$

$$M_s^{(1)}h^e_{Z_1}(x) = [m_{Y_1}(k)U_{Y_1} + m_{Z_1}(k)U_{Z_1}(k) + m_{Y_2}(k)U_{Y_2} + m_{Z_2}(k)U_{Z_2}]e^{ikx},$$

$$M_s^{(2)}h^e_{Y_2}(x) = [m_{Y_1}(k)V_{Y_1} + m_{Z_1}(k)V_{Z_1} + m_{Y_2}(k)V_{Y_2} + m_{Z_2}(k)V_{Z_2}]e^{ikx},$$

and

$$M_s^{(2)}h^e_{Z_2}(x) = [m_{Y_1}(k)W_{Y_1} + m_{Z_1}(k)W_{Z_1} + m_{Y_2}(k)W_{Y_2} + m_{Z_2}(k)W_{Z_2}]e^{ikx}.$$

Thus, Eq. (2) can be expressed in matrix form, this is

$$i\frac{\omega}{\gamma}\,\mathbf{m}(k) = \tilde{\mathbf{A}}\,\mathbf{m}(k), \tag{3}$$

where the transpose of matrix $\mathbf{m}(k)$ is given by $\mathbf{m}^T(k) = [m_{Y_1}(k), m_{Z_1}(k), m_{Y_2}(k), m_{Z_2}(k)]$ and

$$\tilde{\mathbf{A}} = \begin{pmatrix} A^{Y_1}_{Y_1} & A^{Y_1}_{Z_1} & A^{Y_1}_{Y_2} & A^{Y_1}_{Z_2} \\ A^{Z_1}_{Y_1} & A^{Z_1}_{Z_1} & A^{Z_1}_{Y_2} & A^{Z_1}_{Z_2} \\ A^{Y_2}_{Y_1} & A^{Y_2}_{Z_1} & A^{Y_2}_{Y_2} & A^{Y_2}_{Z_2} \\ A^{Z_2}_{Y_1} & A^{Z_2}_{Z_1} & A^{Z_2}_{Y_2} & A^{Z_2}_{Z_2} \end{pmatrix}.$$

The matrix elements are given by

$$A^{Y_1}_\beta = U_\beta - H^{e0}_{X_1}\delta_{\beta,Z_1}, \tag{4a}$$

$$A^{Z_1}_\beta = -T_\beta + H^{e0}_{X_1}\delta_{\beta,Y_1}, \tag{4b}$$

$$A^{Y_2}_\beta = W_\beta - H^{e0}_{X_2}\delta_{\beta,Z_2}, \tag{4c}$$

$$A^{Z_2}_\beta = -V_\beta + H^{e0}_{X_2}\delta_{\beta,Y_2}. \tag{4d}$$

Here $\delta_{\beta,\beta'}$ is the Kronecker delta function and $\beta = Y_1, Z_1, Y_2$ and $Z_2$. Therefore, once the effective fields are derived the elements $T_\beta$, $U_\beta$, $V_\beta$, $W_\beta$ and $H_{X_\nu}^{e0}$ can be readily obtained and hence the matrix $\tilde{\mathbf{A}}$ can be analytically described. By considering Zeeman energy, uniaxial anisotropy with easy axis along $y$, intralayer and interlayer exchange coupling, and dipolar interaction, the elements $T_\beta$, $U_\beta$, $V_\beta$, $W_\beta$ and $H_{X_\nu}^{e0}$ are given by

$$T_{Y_1} = -M_s^{(1)} \sin^2 \varphi_1 \left(1 - \zeta[k, d_1]\right) - M_s^{(1)} [\lambda_{ex}^{(1)}]^2 k^2 + H_u^{(1)} \cos^2 \varphi_1, \tag{5a}$$

$$T_{Y_2} = -M_s^{(1)} \sin \varphi_1 \sin \varphi_2 \frac{|k| d_2}{2} \zeta[k, d_1] \zeta[k, d_2] e^{-|k|s} + \frac{J}{d_1 M_s^{(2)}} \cos(\varphi_2 - \varphi_1),$$

$$T_{Z_1} = 0, \tag{5b}$$

$$T_{Z_2} = i M_s^{(1)} \sin \varphi_1 \frac{k d_2}{2} \zeta[k, d_1] \zeta[k, d_2] e^{-|k|s}; \tag{5c}$$

$$U_{Y_1} = 0, \tag{6a}$$

$$U_{Y_2} = i M_s^{(1)} \sin \varphi_2 \frac{k d_2}{2} \zeta[k, d_1] \zeta[k, d_2] e^{-|k|s}, \tag{6b}$$

$$U_{Z_1} = -M_s^{(1)} \zeta(k, d_1) - M_s^{(1)} [\lambda_{ex}^{(1)}]^2 k^2, \tag{6c}$$

$$U_{Z_2} = M_s^{(1)} \frac{|k| d_2}{2} \zeta[k, d_1] \zeta[k, d_2] e^{-|k|s} + \frac{J}{d_1 M_s^{(2)}}; \tag{6d}$$

$$V_{Y_1} = -M_s^{(2)} \sin \varphi_1 \sin \varphi_2 \frac{|k| d_1}{2} \zeta[k, d_1] \zeta[k, d_2] e^{-|k|s} + \frac{J}{d_2 M_s^{(1)}} \cos(\varphi_1 - \varphi_2), \tag{7a}$$

$$V_{Y_2} = -M_s^{(2)} \sin^2 \varphi_2 \left(1 - \zeta[k, d_2]\right) - M_s^{(2)} [\lambda_{ex}^{(2)}]^2 k^2 + H_u^{(2)} \cos^2 \varphi_2, \tag{7b}$$

$$V_{Z_1} = -i M_s^{(2)} \sin \varphi_2 \frac{k d_1}{2} \zeta[k, d_1] \zeta[k, d_2] e^{-|k|s}, \tag{7c}$$

$$V_{Z_2} = 0, \tag{7d}$$

and

$$W_{Y_1} = -i M_s^{(2)} \sin \varphi_1 \frac{k d_1}{2} \zeta[k, d_1] \zeta[k, d_2] e^{-|k|s}, \tag{8a}$$

$$W_{Y_2} = 0, \tag{8b}$$

$$W_{Z_1} = M_s^{(2)} \frac{|k| d_1}{2} \zeta[k, d_1] \zeta[k, d_2] e^{-|k|s} + \frac{J}{d_2 M_s^{(1)}}, \tag{8c}$$

$$W_{Z_2} = -M_s^{(2)} \zeta[k, d_2] - M_s^{(2)} [\lambda_{ex}^{(2)}]^2 k^2. \tag{8d}$$

Here, we have defined the function

$$\zeta[k, d_\nu] = \frac{\sinh[k d_\nu / 2] e^{-|k| d_\nu / 2}}{k d_\nu / 2}. \tag{9}$$

Also, the $X$-component of the static effective field is given by

$$H_{X_\nu}^{e0} = H\cos(\varphi_h - \varphi_\nu) + H_u^{(\nu)}\sin^2\varphi_\nu + \frac{J}{d_\nu M_s^{(\nu)}}\cos(\varphi_1 - \varphi_2). \tag{10}$$

Explicit expressions for the effective magnetic fields are derived in the following section, where all magnetic parameters are defined.

## EFFECTIVE MAGNETIC FIELDS

### A.  Zeeman field

For the in-plane geometry shown in Fig. 1, the $X_\nu$-component of the external field is

$$H_{X_\nu}^0 = H\cos(\varphi_h - \varphi_\nu), \tag{11}$$

where $H$ is the strength of the dc applied field.

### B.  Uniaxial anisotropy

For simplicity, it is assumed uniaxial anisotropy with easy axis along $y$ for both layers. Thus, the energy density in this case is given by

$$\epsilon_u^{(\nu)} = -\frac{\mu_0 H_u^{(\nu)}}{2 M_s^{(\nu)}} \left[\mathbf{M}^{(\nu)}(x) \cdot \hat{y}\right]^2, \tag{12}$$

where $H_u^{(\nu)} = 2K_u^{(\nu)}/\mu_0 M_s^{(\nu)}$ is the magnitude of the uniaxial anisotropy field, with $K_u^{(\nu)}$ the uniaxial anisotropy constant. Then, $\hat{y} = \cos\varphi_\nu \hat{Y}_\nu + \sin\varphi_\nu \hat{X}_\nu$, and therefore

$$\epsilon_u^{(\nu)} = -\frac{\mu_0 H_u^{(\nu)}}{2 M_s^{(\nu)}} \left[m_{Y_\nu}(x)\cos\varphi_\nu + M_{X_\nu}\sin\varphi_\nu\right]^2. \tag{13}$$

By considering up to second order in the magnetization deviation, the $X$-component and dynamic part of the uniaxial anisotropy field are

$$H_{X_\nu}^{u0} = H_u^{(\nu)}\sin^2\varphi_\nu \tag{14}$$

and

$$h_{Y_\nu}^u(x) = \frac{H_u^{(\nu)}}{M_s^{(\nu)}} m_{Y_\nu}(x)\cos^2\varphi_\nu, \tag{15}$$

respectively.

## C. Intralayer exchange field

The exchange field within layer $\nu$ can be cast in the form

$$\mathbf{H}^{\text{ex}(\nu)}(x) = \left\{ \nabla \cdot [\lambda_{\text{ex}}^{(\nu)}(x)]^2 \nabla \right\} \mathbf{M}^{(\nu)}(x). \tag{16}$$

Here $\lambda_{\text{ex}}^{(\nu)}(x) = \sqrt{2A^{(\nu)}/\mu_0 [M_{\text{s}}^{(\nu)}]^2}$ is the exchange length of the layer $\nu$ and $A^{(\nu)}$ its exchange stiffness constant. Writing the magnetization in the linear regime and taking into account that the exchange length is independent of $x$, the exchange field components are $H_{X_\nu}^{\text{ex}0} = 0$ and

$$h_{Y_\nu, Z_\nu}^{\text{ex}}(x) = -[\lambda_{\text{ex}}^{(\nu)}]^2 k^2 m_{Y_\nu, Z_\nu}(x). \tag{17}$$

## D. Dynamic dipolar fields

### 1. Dynamic dipolar fields due to volumetric magnetic charges

The dynamic dipolar field due to volumetric magnetic charges can be obtained as follows. First, as mentioned before, dynamic components of the magnetization can be written as $m_{Y_\nu, Z_\nu}(x) = m_{Y_\nu, Z_\nu}(k)e^{ikx}$. Then, the volumetric magnetic charge density $\rho^{(\nu)}(\mathbf{r}) = -\nabla_{\mathbf{r}} \cdot \mathbf{M}^{(\nu)}(x) = -dM_x^{(\nu)}(x)/dx$, where $M_x^{(\nu)}(\mathbf{r}) = M_{\text{s}}^{(\nu)} \cos\varphi_\nu - m_{Y_\nu}(x)\sin\varphi_\nu$. Therefore

$$\rho^{(\nu)}(\mathbf{r}) = i m_{Y_\nu}(k)(k \sin\varphi_\nu)e^{ikx}. \tag{18}$$

Here, it has been assumed a uniform distribution of the dynamic magnetization along the thickness, therefore the results presented here should apply to thin films. Then, the magnetostatic potential can be calculated from

$$\phi_{\text{v}}^{(\nu)}(\mathbf{r}) = \frac{1}{4\pi} \int \frac{\rho^{(\nu)}(\mathbf{r}')}{|\mathbf{r} - \mathbf{r}'|} d^3\mathbf{r}', \tag{19}$$

which gives

$$\phi_{\text{v}}^{(\nu)}(\mathbf{r}) = i m_{Y_\nu}(k)(k \sin\varphi_\nu) \frac{1}{4\pi} \int \frac{e^{ikx'}}{|\mathbf{r} - \mathbf{r}'|} d^3\mathbf{r}'. \tag{20}$$

The integral in Eq. (20) can be analytically evaluated inside the film ($\nu$), this is

$$\begin{aligned}\int \frac{e^{ikx'}}{|\mathbf{r} - \mathbf{r}'|} d^3\mathbf{r}' &= 2\pi e^{ikx} \int_{\xi_\nu}^{d_\nu + \xi_\nu} \frac{e^{-|k||z-z'|}}{|k|} dz' \\ &= 2\pi e^{ikx} \left[ \int_{\xi_\nu}^{z} \frac{e^{-|k|(z-z')}}{|k|} dz' + \int_{z}^{d_\nu + \xi_\nu} \frac{e^{-|k|(z'-z)}}{|k|} dz' \right] \\ &= 4\pi \frac{1 - \cosh[k(d_\nu/2 + \xi_\nu - z)]e^{-|k|d_\nu/2}}{k^2} e^{ikx}, \end{aligned} \tag{21}$$

where, according to Fig. 1, $\xi_\nu = (d_{\eta_\nu} + s)\delta_{1,\nu}$ with $\eta_\nu = 2\delta_{1,\nu} + \delta_{\nu,2}$ ($\eta_1 = 2$ and $\eta_2 = 1$). Thus, $\xi_\nu$ is defined in such a way that $\xi_1 = d_2 + s$ and $\xi_2 = 0$. Note that the coordinate $z$ is measured from the bottom of the layer 2. Then, the potential $\phi_v^{(\nu)}(\mathbf{r})$ becomes

$$\phi_v^{(\nu)}(\mathbf{r}) = im_{Y_\nu}(k)\sin\varphi_\nu \frac{1 - \cosh[k(d_\nu/2 + \xi_\nu - z)]e^{-|k|d_\nu/2}}{k}e^{ikx}. \tag{22}$$

The dynamic dipolar field components can be obtained from $h_{x,y,z}^v(\mathbf{r}) = -\partial_{x,y,z}\phi_v^{(\nu)}(\mathbf{r})$. i.e.,

$$h_x^v(\mathbf{r}) = m_{Y_\nu}(k)\sin\varphi_\nu \left(1 - \cosh[k(d_\nu/2 + \xi_\nu - z)]e^{-|k|d_\nu/2}\right)e^{ikx}. \tag{23}$$

$$h_z^v(\mathbf{r}) = -im_{Y_\nu}(k)\sin\varphi_\nu \sinh[k(d_\nu/2 + \xi_\nu - z)]e^{-|k|d_\nu/2}e^{ikx}, \tag{24}$$

and $h_y^v = 0$. Now, the fields averaged along the thickness $d_\nu$ give $h_z^v(x) = 0$ and

$$h_x^v(x) = m_{Y_\nu}(k)\sin\varphi_\nu \left(1 - \zeta[k, d_\nu]\right)e^{ikx}, \tag{25}$$

where $\zeta(k, d_\nu)$ is given in Eq. (9). Now, the $Y$-component of the field is $h_{Y_\nu}^v(x) = -h_x^v(x)\sin\varphi_\nu$, and therefore

$$h_{Y_\nu}^v(x) = -m_{Y_\nu}(k)\sin^2\varphi_\nu \left[1 - \zeta(k, d_\nu)\right]e^{ikx}. \tag{26}$$

On the other hand, the interacting potential $\phi_v^{\text{int}(\nu)}(\mathbf{r})$ is generated by the dynamic magnetization of the layer $(\nu)$, and must be evaluated inside the other layer $(\eta_\nu)$. In this case, the integral in Eq. (20) gives

$$\int \frac{e^{ikx'}}{|\mathbf{r} - \mathbf{r}'|}d^3\mathbf{r}' = 2\pi e^{ikx}\int_{\xi_\nu}^{d_\nu + \xi_\nu}\frac{e^{-|k|(-1)^\nu(z-z')}}{|k|}dz'$$
$$= 4\pi\frac{\sinh[|k|d_\nu/2]}{k^2}e^{(-1)^\nu|k|(d_\nu/2 + \xi_\nu - z)}e^{ikx}.$$

Therefore,

$$\phi_v^{\text{int}(\nu)}(x) = im_{Y_\nu}(k)\sin\varphi_\nu \frac{\sinh[|k|d_\nu/2]}{k}e^{(-1)^\nu|k|(d_\nu/2 + \xi_\nu - z)}e^{ikx}. \tag{27}$$

Now, following the above procedure, and averaging in the respective thickness $\eta_\nu$, the field components projected onto the local coordinate system are

$$h_{Y_{\eta_\nu}}^v = -m_{Y_\nu}(k)\sin\varphi_1\sin\varphi_2\frac{|k|d_\nu}{2}\zeta[k, d_1]\zeta[k, d_2]e^{-|k|s}e^{ikx}, \tag{28}$$

and

$$h_{Z_{\eta_\nu}}^v = (-1)^\nu im_{Y_\nu}(k)\sin\varphi_\nu\frac{kd_\nu}{2}\zeta[k, d_1]\zeta[k, d_2]e^{-|k|s}e^{ikx}. \tag{29}$$

## 2. Dynamic dipolar fields due to superficial magnetic charges

The magnetic potential generated by surface magnetic charges is given by

$$\phi_s^{(\nu)}(\mathbf{r}) = \frac{1}{4\pi} \int \frac{M_z^{(\nu)}(\mathbf{r}')}{|\mathbf{r}-\mathbf{r}'|} dS', \tag{30}$$

where $M_z^{(\nu)}(\mathbf{r}')$ is the normal component of the magnetization. Then,

$$\phi_s^{(\nu)}(\mathbf{r}) = \frac{1}{4\pi} \int_{z'=d_\nu+\xi_\nu} \frac{m_{Z_\nu}(k) e^{ikx'}}{|\mathbf{r}-\mathbf{r}'|} dS' - \frac{1}{4\pi} \int_{z'=\xi_\nu} \frac{m_{Z_\nu}(k) e^{ikx'}}{|\mathbf{r}-\mathbf{r}'|} dS'.$$

Taking into account that

$$\int \frac{e^{ikx'}}{|\mathbf{r}-\mathbf{r}'|} dS' = 2\pi \frac{e^{-|k||z-z'|}}{|k|} e^{ikx},$$

the magnetic potential becomes

$$\phi_s^{(\nu)}(x,z) = m_{Z_\nu}(k) \sinh\left[k\left(z - d_\nu/2 - \xi_\nu\right)\right] \frac{e^{-|k|d_\nu/2}}{k} e^{ikx}. \tag{31}$$

Then, by taking the average along the thickness, it is obtained $h_x^s(x) = 0$ and

$$h_{Z_\nu}^s(x) = -m_{Z_\nu}(k) \zeta[k, d_\nu] e^{ikx}. \tag{32}$$

On the other hand, the interacting potential $\phi_s^{\text{int}(\nu)}(x)$ generated by layer $(\nu)$ and evaluated inside the other magnetic layer $(\eta_\nu)$ is

$$\phi_s^{\text{int}(\nu)}(x,z) = \frac{m_{Z_\nu}(k)}{2} \left[ \left(\frac{e^{-|k|(-1)^\nu(z-z')}}{|k|}\right)_{z'=d_\nu+\xi_\nu} - \left(\frac{e^{-|k|(-1)^\nu(z-z')}}{|k|}\right)_{z'=\xi_\nu} \right] e^{ikx}$$

i.e.,

$$\phi_s^{\text{int}(\nu)}(x,z) = (-1)^\nu m_{Z_\nu}(k) \frac{\sinh[kd_\nu/2]}{k} e^{-(-1)^\nu |k|(z-\xi_\nu-d_\nu/2)} e^{ikx}. \tag{33}$$

Then, the components of the dynamic dipolar fields induced by surface charges are

$$h_x^{s(\nu)}(x,z) = -(-1)^\nu i k m_{Z_\nu}(k) \frac{\sinh[kd_\nu/2]}{k} e^{(-1)^\nu |k|(d_\nu/2+\xi_\nu-z)} e^{ikx} \tag{34}$$

and

$$h_z^{s(\nu)}(x,z) = m_{Z_\nu}(k) \sinh\left[|k|\, d_\nu/2\right] e^{(-1)^\nu |k|(d_\nu/2+\xi_\nu-z)} e^{ikx}. \tag{35}$$

Therefore, the averaged quantities projected onto the axes $Y_{\eta_\nu}$ and $Z_{\eta_\nu}$ are

$$h_{Y_{\eta_\nu}}^s(x) = (-1)^\nu i m_{Z_\nu}(k) \sin\varphi_{\eta_\nu} \frac{kd_\nu}{2} \zeta[k,d_1]\zeta[k,d_2] e^{-|k|s} e^{ikx} \tag{36}$$

and

$$h_{Z_{\eta_\nu}}^s(x) = m_{Z_\nu}(k) \frac{|k|\, d_\nu}{2} \zeta[k,d_1]\zeta[k,d_2] e^{-|k|s} e^{ikx}. \tag{37}$$

### E. Interlayer exchange coupling

The interlayer exchange energy per unit area is given by

$$\epsilon_{\text{int}} = -\frac{J}{M_s^{(\nu)} M_s^{(\eta_\nu)}} \left[ \mathbf{M}^{(\nu)}(\mathbf{r}) \cdot \mathbf{M}^{(\eta_\nu)}(\mathbf{r}) \right], \tag{38}$$

where $J$ is the interlayer exchange coupling parameter between both layers. By taking into account that the magnetization can be written as $\mathbf{M}^{(\nu)}(\mathbf{r}) = \mathbf{M}^{(\nu)}(x) = M_s^{(\nu)} \hat{X}_\nu + m_{Y_\nu}(x) \hat{Y}_\nu + m_{Z_\nu}(x) \hat{Z}_\nu$, Eq. (38) becomes

$$\begin{aligned}\epsilon_{\text{int}} = -\frac{J}{M_s^{(\nu)} M_s^{(\eta_\nu)}} \big[ &M_s^{(\nu)} M_s^{(\eta_\nu)} \cos(\varphi_{\eta_\nu} - \varphi_\nu) - M_s^{(\nu)} m_{Y_{\eta_\nu}}(x) \sin(\varphi_{\eta_\nu} - \varphi_\nu) \\ &+ M_s^{(\eta_\nu)} m_{Y_\nu}(x) \sin(\varphi_{\eta_\nu} - \varphi_\nu) + m_{Y_{\eta_\nu}}(x) m_{Y_\nu}(x) \cos(\varphi_{\eta_\nu} - \varphi_\nu) + m_{Z_\nu}(x) m_{Z_{\eta_\nu}}(x) \big].\end{aligned}$$

Here, it has been used $\hat{Z}_{\eta_\nu} = \hat{Z}_\nu$, $\hat{Y}_{\eta_\nu} = \cos(\varphi_{\eta_\nu} - \varphi_\nu) \hat{Y}_\nu - \sin(\varphi_{\eta_\nu} - \varphi_\nu) \hat{X}_\nu$ and $\hat{X}_{\eta_\nu} = \cos(\varphi_{\eta_\nu} - \varphi_\nu) \hat{X}_\nu + \sin(\varphi_{\eta_\nu} - \varphi_\nu) \hat{Y}_\nu$. Then, the $X$-component of the static interlayer exchange field is given by

$$H_{X_\nu}^{\text{int}0} = \frac{J}{d_\nu M_s^{(\nu)}} \cos(\varphi_{\eta_\nu} - \varphi_\nu), \tag{39}$$

while the $Z_\nu$- and $Y_\nu$-components of the dynamic interlayer exchange field are

$$h_{Z_\nu}^{\text{int}}(x) = \frac{J}{d_\nu M_s^{(\nu)} M_s^{(\eta_\nu)}} m_{Z_{\eta_\nu}}(k) e^{ikx} \tag{40}$$

and

$$h_{Y_\nu}^{\text{int}}(x) = \frac{J}{d_\nu M_s^{(\nu)} M_s^{(\eta_\nu)}} m_{Y_{\eta_\nu}}(k) \cos(\varphi_{\eta_\nu} - \varphi_\nu) e^{ikx}. \tag{41}$$

Now, it is straightforward to include all the fields into Eq. (4), in such a way that the elements of matrix $\tilde{\mathbf{A}}$ in Eq. (3) can be readily obtained.

Note that the fields calculated here are valid for a bilayer system, where both dipolar and exchange interaction is assumed between FM layers. Nevertheless, the dynamic dipolar fields derived in section I D can be readily generalized for a multilayer structure, since the calculations allow determining the dynamic stray fields above and below of one FM layer and therefore, the interaction of more than two FM layers is feasible. In this sense, if the upper and bottom layer split each one into $N$ sublayers, where all sublayers are coupled via dipolar interaction whilst the interlayer exchange interaction is present only between nearest neighbors, the theoretical description is able to take into account the dynamics of thicker FM films. In other words, this subdivision takes into account the variation of the dynamic magnetization along the thickness and therefore allows to describe the dynamic of films of around 40 nm of thickness, such as the samples described in the manuscript. Note that when the system is subdivided into $N$ sublayers, the exchange interaction between them is of intralayer kind, defined by a constant $J^{(\nu)}_{\text{intra}}$. Of course, this constant $J^{(\nu)}_{\text{intra}}$ must be related with the exchange constant $A^{(\nu)}$ defined in I C. This relation is shown in the following section.

## F. Continuous limit for Intralayer exchange interaction

When layer $\nu$ is subdivided into $N$ sublayers, the exchange energy per unit area between the sublayer $i$ and $i+1$ is

$$\epsilon^{\text{ex}}_{\text{intra}} = -\sum_i \frac{J^{(\nu)}_{\text{intra}}}{[M^{(\nu)}_{\text{s}}]^2} \left[ \mathbf{M}^{(\nu_i)}(z_i) \cdot \mathbf{M}^{(\nu_{i+1})}(z_i + a) \right]. \tag{42}$$

This energy considers the exchange interaction between neighbors magnetic moments separated by a distance $a$, which corresponds to the thickness of each sublayer $i$. Since the FM layer can be subdivided into many sublayers, a slow variation of the magnetization along the normal $z$-axis can be assumed, and therefore, the expression (42) takes the form:

$$\epsilon^{\text{ex}}_{\text{intra}} = -\sum_i \frac{J^{(\nu)}_{\text{intra}}}{[M^{(\nu)}_{\text{s}}]^2} \left[ \mathbf{M}^{(\nu_i)}(z_i) \cdot \left( \mathbf{M}^{(\nu_i)}(z_i) + a \frac{\partial \mathbf{M}^{(\nu_i)}(z_i)}{\partial z} + \frac{a^2}{2} \frac{\partial^2 \mathbf{M}^{(\nu_i)}(z_i)}{\partial z^2} \right) \right] \tag{43}$$

Under the normalization condition for the magnetization of each sublayer, the last expression can be written as

$$\epsilon^{\text{ex}}_{\text{intra}} = -\sum_i \frac{J^{(\nu)}_{\text{intra}}}{[M^{(\nu)}_{\text{s}}]^2} \left[ \mathbf{M}^{(\nu_i)}(z_i) \cdot \left( \frac{a^2}{2} \frac{\partial^2 \mathbf{M}^{(\nu_i)}(z_i)}{\partial z^2} \right) \right], \tag{44}$$

where the constant term has been omitted. Now, expressing the second derivatives in terms of the first derivatives, Eq. (44) becomes

$$\begin{aligned} \epsilon^{\text{ex}}_{\text{intra}} &= \sum_i \frac{J^{(\nu)}_{\text{intra}} a^2}{2[M^{(\nu)}_{\text{s}}]^2} \left( \frac{\partial \mathbf{m}(z_i)}{\partial z} \right)^2 \\ &\approx \frac{J^{(\nu)}_{\text{intra}} a}{2} \int \left( \frac{\partial \hat{\mathbf{m}}(z_i)}{\partial z} \right)^2 dz \\ &= A^{(\nu)} \int \left( \frac{\partial \hat{\mathbf{m}}(z_i)}{\partial z} \right)^2 dz, \end{aligned}$$

where $\hat{\mathbf{m}}(z_i) = \mathbf{m}(z_i)/M_{\mathrm{s}}^{(\nu)}$ is the normalized dynamic magnetization. Therefore $J_{\mathrm{intra}}^{(\nu)} = 2A^{(\nu)}/a$.